# Slow relaxation in a one-dimensional rational assembly of antiferromagnetically-coupled [$Mn_4$] single-molecule-magnets

*Lollita Lecren,[†] Olivier Roubeau,[†] Claude Coulon,[†,]* Yang-Guang Li,[†,ƒ] Xavier Le Goff,[†] Wolfgang Wernsdorfer,[‡] Hitoshi Miyasaka,[§] and Rodolphe Clérac[†,]**

Centre de Recherche Paul Pascal, CNRS UPR-8641, 115 av. Albert Schweitzer, 33600 Pessac, France; Laboratoire Louis Néel, CNRS, BP 166, 25 av. des Martyrs, 38042 Grenoble, France; Department of Chemistry, Graduate School of Science, Tokyo Metropolitan University, Minami-ohsawa 1-1, Hachioji, Tokyo 192-0397, PRESTO and CREST, Japan Science and Technology Agency, 4-1-8 Honcho, Kawaguchi, Saitama 332-0012, Japan.

E-mail: clerac@crpp-bordeaux.cnrs.fr; coulon@crpp-bordeaux.cnrs.fr

**Running title:** Slow relaxation in a 1-D rational assembly of antiferromagnetically-coupled SMM

---

## Abstract:

Four discrete $Mn^{III}/Mn^{II}$ tetra-nuclear complexes with double-cuboidal core, $[Mn_4(hmp)_6(CH_3CN)_2(H_2O)_4](ClO_4)_4 \cdot 2CH_3CN$ (**1**), $[Mn_4(hmp)_6(H_2O)_4](ClO_4)_4 \cdot 2H_2O$ (**2**), $[Mn_4(hmp)_6(H_2O)_2(NO_3)_2](ClO_4)_2 \cdot 4H_2O$ (**3**) and $[Mn_4(hmp)_6(Hhmp)_2](ClO_4)_4 \cdot 2CH_3CN$ (**4**) were synthesized by reaction of Hhmp (2-hydroxymethylpyridine) with $Mn(ClO_4)_2 \cdot 4H_2O$ in presence of tetraethylammonium hydroxide and subsequent addition of $NaNO_3$ (**3**) or an excess of Hhmp (**4**). dc magnetic measurements show that both $Mn^{2+}$—$Mn^{3+}$ and $Mn^{3+}$—$Mn^{3+}$ magnetic interactions are ferromagnetic in **1**-**3** leading to an $S_T = 9$ ground state for the $Mn_4$ unit. Furthermore, these complexes are Single-Molecule Magnets (SMMs) clearly showing both thermally activated and ground state tunneling regimes. Slight changes in the $[Mn_4]$ core geometry result in an $S_T = 1$ ground state in **4**. A one-dimensional assembly of $[Mn_4]$ units, catena-$\{[Mn_4(hmp)_6(N_3)_2](ClO_4)_2\}$ (**5**), was obtained in the same synthetic conditions with the subsequent addition of $NaN_3$. Double chair-like $N_3^-$ bridges connect identical $[Mn_4]$ units into a chain arrangement. This material behaves as an Ising assembly of $S_T = 9$ tetramers weakly antiferromagnetically coupled. Slow relaxation of the magnetization is observed at low temperature for the first time in an antiferromagnetic chain, following an activated behavior with $\Delta_\tau/k_B = 47$ K and $\tau_0 = 7\times10^{-11}$ s. The observation of this original thermally activated relaxation process is induced by finite-size effects and in particular by the non-compensation of spins in segments of odd-number units. Generalizing the known theories on the dynamic properties of poly-disperse finite segments of antiferromagnetically coupled Ising spins, the theoretical expression of the characteristic energy gaps $\Delta_\xi$ and $\Delta_\tau$ were estimated and successfully compared to the experimental values.

## Introduction

Slow-relaxing magnetic nano-systems are thought of being able to bring technological breakthroughs to information storage and optical applications.[1] Aside nano-particles of classical magnets,[2] poly-nuclear transition metal complexes have attracted much attention in this field, from both synthetic and theoretical viewpoints.[3] Families of such systems, so-called single-molecule magnets (SMMs), have thus been obtained using a variety of transition metal ions. The slow reversal of their magnetization at low temperature arises from the combined effect of a high-spin ground state and uni-axial anisotropy resulting in an energy barrier between spin-up and spin-down states. Depending on the temperature, this relaxation obeys two processes. At high temperatures, the relaxation time, $\tau$, is thermally activated and the theoretical energy barrier $\Delta$ is equal to $|D|S_T^2$ for integer spin and $|D|(S_T^2\text{-}1/4)$ for half-integer spin. At very low temperatures, quantum tunneling of the magnetization (*QTM*) governed by the transverse anisotropy ($E$), becomes the fastest pathway of relaxation. Experimentally a crossover occurs between these two regimes called thermally assisted *QTM*. In this intermediate range of temperature, the thermal barrier is "short-cut" by quantum tunneling and an effective barrier, $\Delta_{eff}$, is found smaller than $\Delta$. In many SMM systems, this regime is the only one seen experimentally before that $\tau$ becomes temperature independent. Another kind of molecular-based coordination materials have been shown more recently to present slow-relaxation of the magnetization,[4a] namely the so-called single-chain magnets (SCMs).[5] In these materials, the slow relaxation of magnetization is not solely the consequence of a high-spin ground state and the uni-axial anisotropy seen by each spin along the chain, but depends also on magnetic correlations. These compounds are one-dimensional assemblies of spins (metal ions, organic radicals or metal-ion clusters) that can be coupled ferro-[5,6,7b] or antiferromagnetically.[4] The latter case corresponds so far only to systems with a non-compensation of spins along the chain (ferrimagnetic or canted-antiferromagnetic arrangements). Their design relies mostly on the bottom-up approach, in which building blocks presenting labile coordinating sites and relevant magnetic properties (uni-axial anisotropy and high spin ground state) are assembled into one-dimension. This step-by-step synthetic strategy has been recently extended to SMM building-blocks that have been used to design SCM materials.[6g,7] Among the known SMMs interesting for this strategy, D. N. Hendrickson and G. Christou have reported in 2001 a tetranuclear mixed-valence SMM: $[Mn_4(hmp)_6Br_2(H_2O)_2]Br_2\cdot4H_2O$,[8a,8c] and more recently two other derivatives (hmp stands for the anion of Hhmp, 2-hydroxymethylpyridine).[8b] This type of complex possesses a rhombic-core $[Mn_4]$ of a rather small dimension in comparison with its $S_T = 9$ ground state, and its terminal ligands on two opposite sides are always either coordinating anions and/or solvent molecules. In the past few years, our group has been interested in the synthesis and study of more specimens of this family,[9] with the goal of linking them in a controlled manner. Using this strategy, we have shown recently that a ferrimagnetic order can be observed in a material where $[Mn_4]$ SMM are connected through $[Mn(N(CN)_2)_6]^{4-}$ units in a three-dimensional architecture.[10] In 2005, D. N. Hendrickson et al showed that one-dimensional arrangements can also be achieved serendipitously using $[Mn_4]$ building blocks.[11] Herein, we demonstrate that quite generally external ligands on $[Mn_4]$ are indeed easily exchangeable, and that the use of coordinating bridging anions allows the controlled synthesis of a $[Mn_4]$ one-dimensional assembly. The magnetic properties of the resulting $[Mn_4]$ building blocks and the chain compound are presented and discussed in details. In particular, we show for the first time that finite-size effects allow the detection of magnetization slow relaxation in a regular antiferromagnetic chain.

## Experimental Section

***General Procedures and Materials.*** All manipulations were carried out under aerobic conditions using commercial grade solvents. Manganese perchlorate hexahydrate (ABCR), 2-hydroxymethyl-pyridine (Hhmp, Aldrich), tetraethylammonium hydroxide (TEAOH, Aldrich), sodium azide (Acros) and sodium nitrate (Aldrich) were all used as received without further purification. *Caution! Perchlorate salts are potentially explosive and should only be handled in small quantities.*

***Synthesis.***

**$[Mn_4(hmp)_6(CH_3CN)_2(H_2O)_4](ClO_4)_4\cdot2CH_3CN$ (1)**. Complex **1** was obtained as previously described[9a] from $Mn(ClO_4)_2\cdot6H_2O$ and 2-hydroxymethylpyridine (Hhmp) in acetonitrile.

**$[Mn_4(hmp)_6(H_2O)_4](ClO_4)_4{\cdot}2H_2O$** (**2**). This compounds slowly forms (ca. 2-3 months) as red crystals from crystals of **1** left in their mother liquor. It is also obtained by putting crystals of **1** in an acetonitrile/water mixture. The crystals of **2** are highly sensitive to solvent loss. Bulk magnetization for **2** was obtained on a "wet" sample, and low temperature single-crystal magnetization measurements (*μ*-SQUID technique) could not be performed as a result of this sensitivity. The low quality of the single-crystal X-ray data results also from the rapid loss of solvent molecules.

**$[Mn_4(hmp)_6(H_2O)_2(NO_3)_2](ClO_4)_2{\cdot}4H_2O$ (3).** $Mn(ClO_4)_2{\cdot}6H_2O$ (0.500 g, 1.38 mmol) and Hhmp (0.378 g, 3.45 mmol) were dissolved in 20 mL of $CH_3CN$ with stirring. To this solution, 0.530 g (0.72 mmol) of a 20 wt % water solution of TEAOH were added dropwise. Then, solid $NaNO_3$ (0.049 g, 0.58 mmol) was added to the above solution and the red-brown solution was stirred for 1 h at room temperature. After filtration, the solution was kept undisturbed in a flask for slow evaporation. Red-brown prismatic crystals of **3** were isolated within a day to a week. The crystals were collected by filtration, washed with $CH_3CN$, and deposited in vacuo. Yield: 51%. Anal. Calc. for $[Mn_4(hmp)_6(H_2O)_2(NO_3)_2](ClO_4)_2{\cdot}4H_2O$ ($C_{36}H_{50}N_8O_{26}Cl_2Mn_4$): C, 33.23; H, 3.88; N, 8.62%; Found: C, 33.62; H, 3.86; N, 8.31 %. Selected IR data (KBr, $cm^{-1}$) 3411 (s), 2908 (w), 2845 (m), 1604 (s), 1560 (s), 1476 (s), 1433 (s), 1386 (s), 1312 (s), 1288 (s), 1222 (m), 1144 (m), 1114 (s), 1084 (s), 1057 (s), 1041 (s), 822 (m), 755 (s), 715 (m), 672 (s), 621 (s), 568 (s), 531 (m), 484 (w), 410 (m).

**$[Mn_4(hmp)_6(Hhmp)_2](ClO_4)_4{\cdot}2CH_3CN$ (4).** 0.150 mg of $[Mn_4(hmp)_6(CH_3CN)_2(H_2O)_4](ClO_4)_4{\cdot}2CH_3CN$ (**1**) were dissolved in 10 mL of acetonitrile. To this solution transferred in a test tube (2 cm in diameter), a 20 ml diethylether solution containing 44 mg (0.40 mmol) of Hhmp was gently added. The tube was then sealed and kept undisturbed. After ca. one week, a kinetic product formed as hexagonal dark-red crystals, identified by x-ray diffraction as a heptameric mixed valence Mn complex. Keeping the tube undisturbed without collecting the crystals of the kinetic species, a thermodynamic product was obtained after ca. one month as dark-pink crystals of **4**. These latter crystals were filtered, washed with acetonitrile and dried in air. Yield: 41 %. Elemental analysis indicates the loss of the solvent molecules. Anal. Calc. for $[Mn_4(hmp)_6(Hhmp)_2](ClO_4)_4$: C 38.87 ; H 3.40 ; N 7.56 % ; Found : C 39.03 ; H 3.50 ; N 8.05 %. Selected IR data (KBr, $cm^{-1}$); 3385(s),

2845(w), 1605(s), 1569(m), 1485(s), 1431(s), 1362(m), 1288(m), 1218(w), 1149(s), 1120(s), 1080(s), 1041(s), 833(w), 764(s), 725(w), 665(w), 626(s), 567(s), 530(w).

**$\{[Mn_4(hmp)_6(N_3)_2](ClO_4)_2\}_\infty$ (5)**. After dissolution of 500 mg (1.38 mmol) of $Mn(ClO_4)_2{\cdot}6H_2O$ in 10 mL of acetonitrile, 335 μL (0,378 g, 3.45 mmol) of Hhmp were added to the solution that took immediately a slight pinkish color. After 1 min stirring, 510 μL of a 20 wt% aqueous solution of TEAOH (530 mg, 0.72 mmol) were added, the solution turning dark red. After 1 min stirring, 76 mg of $NaN_3$ (1.18 mmol) were added and the solution was stirred for one hour in a closed beaker. After filtration, 5 ml of the filtrate were put in a 150 ml beaker that was sealed in a flask containing diethylether. Upon slow diffusion of ether into the unperturbed solution, cubic salmon crystals started to appear after one week. The crystals were filtrated after 10 days, washed with an acetonitrile/toluene mixture (5/3) and dried in air. Although numerous attempts to grow large single-crystal of **5** for traditional magnetic measurements, only tiny single-crystals (typically: $0.3 \times 0.3 \times 0.2$ mm$^3$) allowing X-ray single-crystal diffraction and $\mu$-SQUID measurements have been obtained. Yield: 79 %. Anal. Calc. for $[Mn_4(hmp)_6(N_3)_2](ClO_4)_2$: C 37.55 ; H 3.15 ; N 14.60 % ; Found : C 37.34 ; H 2.50 ; N 13.87 %. Selected IR data (KBr, cm$^{-1}$): 3380 (br), 3110-3030 (w), 2920 (w), 2840 (m), 2095 (s), 1600 (s), 1567 (w), 1488 (w), 1437 (m), 1366 (w), 1280 (m), 1160-1010 (s), 765 (m), 673 (w), 580 (m).

***Physical Measurements.*** Elemental analyses (C, H, and N) were measured by Service Central d'Analyse in CNRS. IR spectra were recorded in the range 400-4000 cm$^{-1}$ on a Nicolet 750 Magna-IR spectrometer using KBr pellets. Magnetic susceptibility measurements were obtained with the use of a Quantum Design SQUID magnetometer (MPMS-XL). dc measurements were conducted from 1.8 to 300 K and from –70 kOe to 70 kOe. ac measurements were performed at frequencies ranging from 0.1 Hz to 1500 Hz with an ac field amplitude of 3 Oe and no dc field applied. The measurements were performed on finely ground polycrystalline samples. Experimental data were corrected for the sample holder and for the diamagnetic contribution of the samples calculated from Pascal constants.[12] Magnetization measurements on single crystals were performed with an array of $\mu$-SQUIDs.[13] This magnetometer works in the temperature range of 0.04 to ~ 7 K and in fields of up to 1.4 T with sweeping rates as high as 10 T/s, along with a field stability of microtesla. The time resolution is approximately 1 ms. The field can be applied in any direction of the μ-SQUID plane with precision much better than 0.1° by separately driving three orthogonal coils. In order to ensure good thermalization, the single crystals were fixed with Apiezon grease.

**Crystallography:** X-ray crystallographic data were collected on a Nonius Kappa CCD diffractometer with graphite-monochromated Mo Kα radiation (λ = 0.71073 Å) at 150(2) K. Suitable crystals were affixed to the end of a glass fiber using silicone grease and transferred to the goniostat. DENZO-SMN[14] was used for data integration and SCALEPACK[14] corrected data for Lorentz-polarisation effects. The structures were solved by direct methods and refined by a full-matrix least-squares method on $F^2$ using the SHELXTL crystallographic software package.[15] Crystal data for **2**-**5** are gathered in Table 1. Note that the loss of solvent molecules in **2** leads quickly to a poor crystallinity and account for the lower quality of the collected data.

## Results and Discussion

### *Synthesis*

Tetra-nuclear mixed-valence $[Mn^{III}_2Mn^{II}_2]$ complexes with a rhombic core can be isolated easily from Mn(II) salts in combination with a chelating-bridging ligand such as Hhmp and an organic base.[8,9] Indeed a number of related compounds have been reported with hmp,[8a-c,9] 2,6-dihydroxymethylpyridine[8c,16a] and triethanolamine.[16b] When a non-coordinating anion is used, terminal coordination sites on the outer Mn(II) ions are occupied by solvent molecules. This is the case in the perchlorate complex $[Mn_4(hmp)_6(CH_3CN)_2(H_2O)_4](ClO_4)_4{\cdot}2CH_3CN$ (**1**), in which terminal positions are occupied by acetonitrile and water molecules.[9a] The $[Mn^{III}_2Mn^{II}_2]$ core forms probably immediately after addition of the precursors as seen by the instantaneous dark red color of the solution. In suitable crystallization conditions, this fast kinetic leads within hours to single-crystals of **1**.[9a] On the other hand, **1** is highly soluble in acetonitrile, and therefore post-synthetic modifications can be envisaged. For instance, complex **1** in $H_2O/CH_3CN$ solution or kept in its mother solution slowly loses its terminal and lattice acetonitrile molecules, forming the thermodynamic product $[Mn_4(hmp)_6(H_2O)_4](ClO_4)_4{\cdot}2H_2O$

(**2**). When the sodium salt of $NO_3^-$ is added to a solution containing **1**, the complex $[Mn_4(hmp)_6(H_2O)_2(NO_3)_2](ClO_4)_2{\cdot}4H_2O$ (**3**) with terminal coordinated nitrato anions is obtained in high yield. Although the actual core of the cluster remains identical, the external coordination sites on Mn(II) ions are labile and thus exchangeable to accept many different anions and/or solvent molecules. This coordination ability gives to these complexes a high flexibility that can be used to build new magnetic units or architectures. Indeed if an excess of Hhmp is added to a solution containing **1**, a new $[Mn_4]$ complex, $[Mn_4(hmp)_6(Hhmp)_2](ClO_4)_4{\cdot}2CH_3CN$ (**4**), is obtained but the terminal positions are this time occupied by protonated Hhmp ligands. Therefore, we have further investigated the versatility of this system to connect $[Mn_4]$ units in a controlled manner, into an extended 1-D structure. For this purpose, we have used the azide anion, a well-known bridging coordinating ligand, that is added as its sodium salt to an acetonitrile solution containing complex **1** (either re-dissolved after isolation or simply at the end of its synthetic procedure). After few days of ether slow diffusion, salmon crystals of $\{[Mn_4(hmp)_6(N_3)_2](ClO_4)_2\}_\infty$ (**5**) form. The stronger and less labile $N_3^-$ anion is able to replace the terminal water/acetonitrile molecules, and when a slight excess of azide is added, the chain compound **5** is formed quantitatively. At this point, it is important to notice that the crystallization process of **5** is slow and the size of the single-crystals is always very limited (typically: $0.3 \times 0.3 \times 0.2$ mm$^3$). If this size of crystal is perfectly suitable for single-crystal X-ray diffraction or $\mu$-SQUID measurements, this is not the case for single-crystal susceptibility measurements using our regular SQUID magnetometer. Therefore, we have extensively investigated the crystallization conditions of **5** in order to obtain large single-crystals. Unfortunately, all attempts were unsuccessful as if the crystal growth was systematically stopped at a critical size. This observation goes in line with the presence of intrinsic defects in the crystals that will be detected by magnetic measurements.

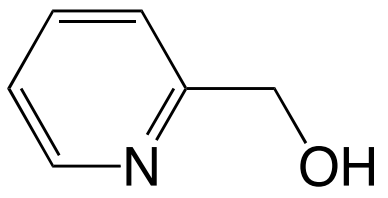

***Scheme 1.*** Hhmp ligand (2-hydroxymethylpyridine).

**Table 1.** Crystallographic data and refinement for compounds **2-5**.

| | **2** | **3** | **4** | **5** |
|---|---|---|---|---|
| Formula | $C_{36}H_{38}Cl_4Mn_4N_6O_{30}$ | $C_{36}H_{50}Cl_2Mn_4N_8O_{26}$ | $Mn_4C_{52}H_{56}N_{10}O_{24}Cl_4$ | $C_{36}H_{36}Cl_2Mn_4N_{12}O_{14}$ |
| Formula Weight | 1396.28 | 1301.50 | 1566.63 | 1151.43 |
| Crystal System | Monoclinic | Triclinic | Monoclinic | Monoclinic |
| Space group | *C*2/c (No. 15) | *P*–1 (No. 2) | $P2_1$/n (No. 14) | $P2_1$/n (No. 14) |
| *a* (Å) | 27.751(6) | 10.084(2) | 14.280(3) | 10.530(2) |
| *b* (Å) | 14.957(3) | 11.692(2) | 15.327(3) | 16.060(3) |
| *c* (Å) | 15.518(3) | 12.112(2) | 14.611(3) | 13.260(3) |
| $\alpha$ (°) | 90 | 78.20(3) | 90 | 90 |
| $\beta$ (°) | 113.80(3) | 84.40(3) | 100.33(3) | 102.79(3) |
| $\gamma$ (°) | 90 | 67.70(3) | 90 | 90 |
| *V* (Å$^3$) | 5894(2) | 1293.0(4) | 3146(1) | 2186.8(8) |
| Z | 2 | 1 | 2 | 4 |
| $D_{calc}$ (g/cm$^3$) | 1.536 | 1.671 | 1.654 | 1.749 |
| *F*(000) | 2744 | 662 | 1596 | 1164 |
| Crystal Size (mm) | 0.4 × 0.2 × 0.2 | 0.56 × 0.38 × 0.34 | 0.42 × 0.36 × 0.29 | 0.3 × 0.3 × 0.2 |
| T (K) | 150(1) | 150(1) | 150(1) | 150(1) |
| θ range for data collection (°) | 1.9–28.1 | 3.40–26.7 | 2.27–30.3 | 2.0–27.4 |
| Index range | $-32 \leq h \leq 36$<br>$-18 \leq k \leq 19$<br>$-19 \leq l \leq 20$ | $-11 \leq h \leq 12$<br>$-14 \leq k \leq 13$<br>$-15 \leq l \leq 13$ | $-20 \leq h \leq 20$<br>$-17 \leq k \leq 21$<br>$-20 \leq l \leq 20$ | $-13 \leq h \leq 13$<br>$-20 \leq k \leq 20$<br>$-17 \leq l \leq 17$ |
| Reflections collected | 19286 | 8569 | 22776 | 33778 |
| Independent reflections | 6864 ($R_{int}$ = 0.121) | 5385 ($R_{int}$ = 0.0742) | 9156 ($R_{int}$ = 0.0876) | 4963 ($R_{int}$ = 0.154) |
| Observed reflections ($I > 2\sigma$(I)) | 2813 | 2948 | 4651 | 2781 |
| Data, restrains, parameters | 6864, 0, 349 | 5385, 24, 372 | 9156, 31, 438 | 4963, 0, 325 |
| $R_1$[a] (observed) | 0.1265 | 0.0681 | 0.0545 | 0.0540 |
| $wR_2$[b] (observed) | 0.4166 | 0.1574 | 0.1186 | 0.1814 |
| *S*[c] | 1.033 | 0.984 | 0.956 | 1.05 |
| Largest diff. peak, hole in eÅ$^{-3}$ | 0.773, –0.900 | 0.466, –0.561 | 0.513, –0.55 | 0.717, –1.268 |

[a] $R_1 = \Sigma||F_o|-|F_c||/\Sigma|F_o|$ ; [b] $wR_2 = [\Sigma w(F_o^2-F_c^2)^2/\Sigma wF_o^4]^{1/2}$ ; $w = 1/\sigma^2(|F_o|)$; [c] Goodness-of-fit $S = [\Sigma w(F_o^2-F_c^2)^2/(n-p)]^{1/2}$, where *n* is the number of reflections and *p* is the number of parameters.

### *Structural Description*

Crystal data and structure refinement details for compound **2-5** are gathered in Table 1. A comparison of relevant structural parameters of the $[Mn_4]$ core in compounds **1-5** is given in Table 2. The tetra-nuclear complex **2** crystallizes in the monoclinic space group *C*2/c with a crystal structure very similar to the recently reported compounds: $[Mn_4(hmp)_6(CH_3CN)_2(H_2O)_4](ClO_4)_4{\cdot}2CH_3CN$ (**1**) and $[Mn_4(hmp)_6(H_2O)_2(NO_3)_2](NO_3)_2{\bullet}2.5H_2O$.[9] The core of **1** and **2** consists of a centro-symmetrical $[Mn_4(hmp)_6(CH_3CN)_{2-x}(H_2O)_4]^{4+}$ (x = 0 and 2 respectively for **1** and **2**) cation, where Mn(2) and Mn(2A) are $Mn^{3+}$, Mn(1) and Mn(1A) are $Mn^{2+}$ and all the hmp are deprotonated. Mn ions are arranged in a double-cuboidal fashion where central $Mn^{3+}$ centers are hexa-coordinated by two pyridine N atoms and four bridging hmp O atoms (see Figure S1). While $Mn^{2+}$ are hepta-coordinated in **1**, the loss of acetonitrile molecules results in a hexa-coordination in **2**. These core geometry and coordination modes have already been observed in $[Mn_4(hmp)_6(CH_3CN)_2(NO_3)_2](ClO_4)_2{\cdot}2CH_3CN$, $[Mn_4(hmp)_6(NO_3)_4]{\cdot}CH_3CN$ and $[Mn_4(hmp)_6Br_2(H_2O)_2]Br_2{\cdot}4H_2O$.[8a-b] $Mn^{3+}$ ions are in a distorted octahedral geometry revealing the expected Jahn-Teller distortion along the axial positions (N(2)–Mn(2) = 2.1902(17) Å and O(1) –Mn(2) = 2.2436(14) Å). Taking one $[Mn_4]$ unit, Jahn-Teller axes are parallel to each other. However in the crystal structure there are two complex orientations, and thus two Jahn-Teller directions are found with an angle of about 26.5° in **1** and 26° in **2**. $ClO_4^-$ anions (disordered) and solvent molecules ($CH_3CN$ and $H_2O$ respectively for **1** and **2**) reside in the inter-space of the 3-D network and interact with the adjacent $[Mn_4]$ clusters through weak hydrogen-bonds. No direct hydrogen bonding or π-π interactions between $[Mn_4]$ units are present in the packing.

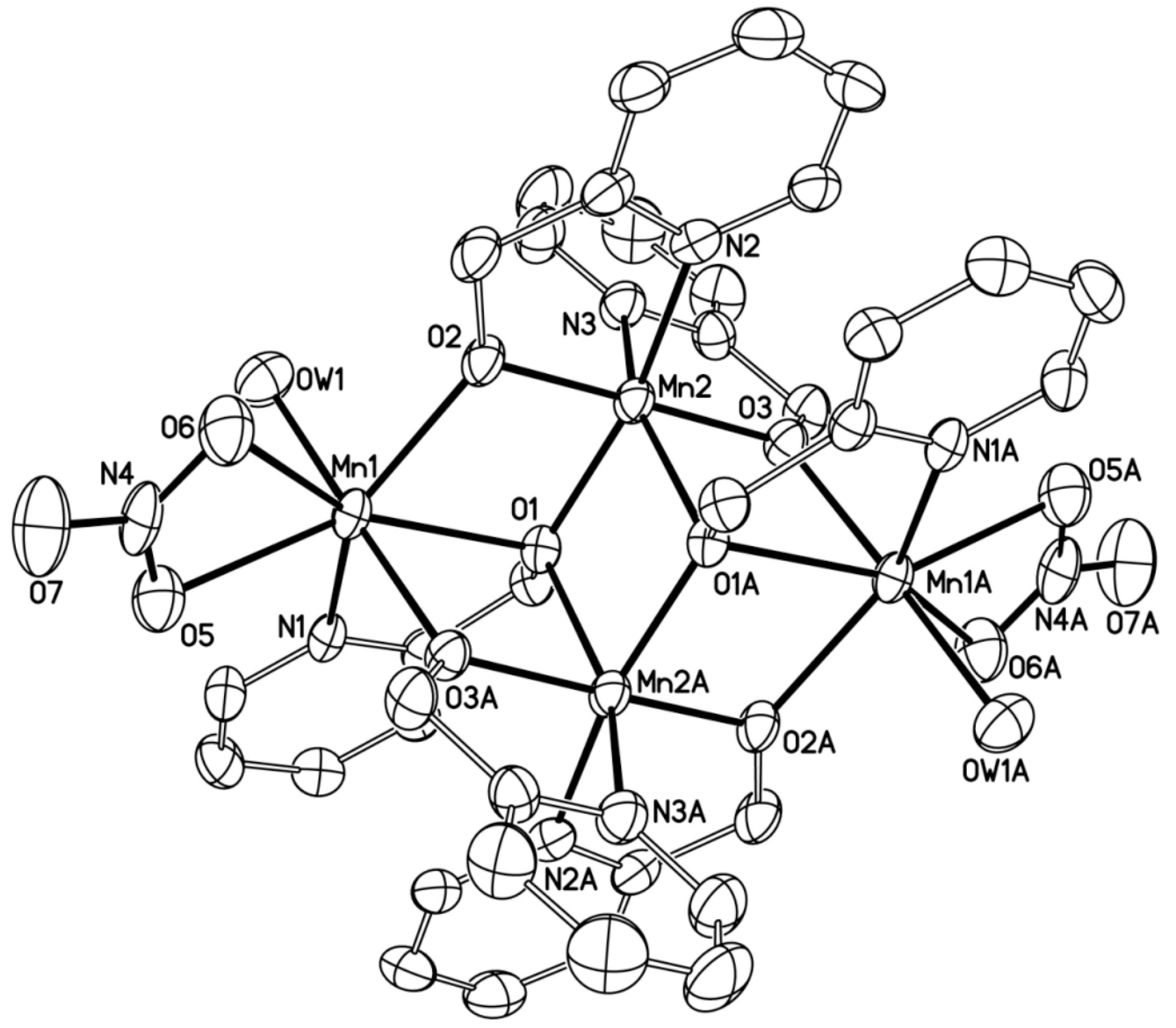

***Figure 1***. ORTEP representation of the cationic parts in **3** with thermal ellipsoids at 30 % probability. H atoms are omitted for clarity.

The nitrate compound **3** crystallizes in the triclinic space group *P*-1 and presents a manganese tetra-nuclear cation $[Mn_4(hmp)_6(H_2O)_2(NO_3)_2]^{2+}$ similar to **1** and **2** (Figure 1). The outer Mn(1) sites are hepta-coordinated as in **1** with a coordination sphere completed by a bidentate $NO_3^-$ ligand. The oxidation states of the two Mn sites, Mn(1) and Mn(2) ions (or Mn(3) and Mn(4) ions), have been assigned as divalent and trivalent respectively. As expected for $Mn^{III}$ ions, Mn(2) (and Mn(4)) shows a Jahn-Teller elongation axis corresponding to Mn(2)–N (ranging from 2.17 to 2.21 Å) and Mn(2)–O (ranging from 2.16 to 2.29 Å) bonds longer than equatorial ones (ca. 1.8 to 2.0 Å). The Jahn-Teller axes are slightly bent (N-Mn-O angle is 159.24(15) °). In the crystal structure, the $[Mn_4]$ complexes possess all the same orientation and this dication unit is charge-balanced by two $ClO_4^-$ counter-anions. Four $H_2O$ solvent molecules fill the void spaces of the unit cell and prevent direct hydrogen bonding or π-π interactions between $[Mn_4]$ units in the packing.

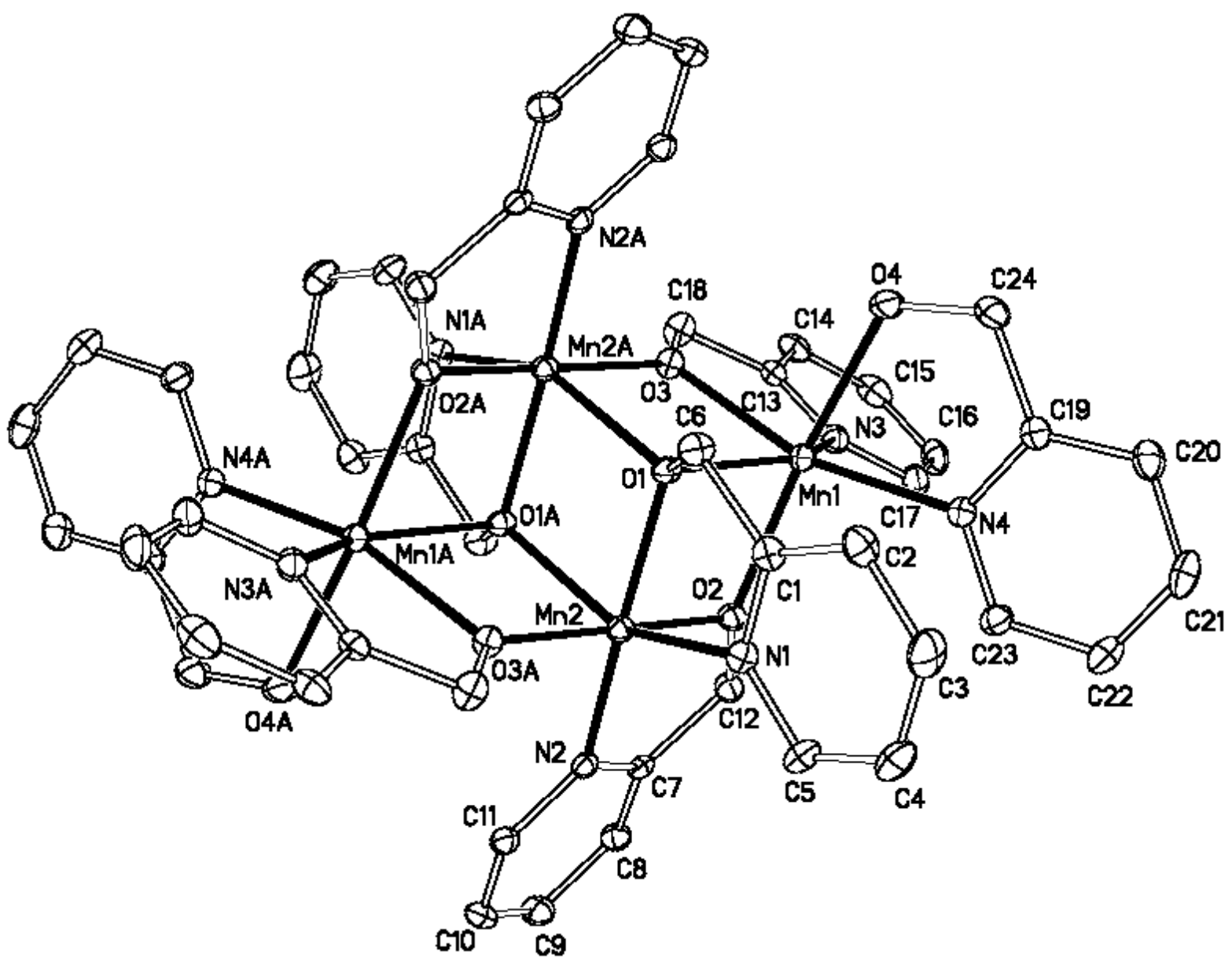

***Figure 2***. ORTEP representation of the cationic part in **4** with thermal ellipsoids at 30 % probability. H atoms are omitted for clarity.

$[Mn_4(hmp)_6(Hhmp)_2](ClO_4)_4{\cdot}2CH_3CN$ (**4**) crystallizes in the monoclinic space group $P2_1/n$ and presents a similar centrosymmetric rhombic $Mn_4O_6$ core as in **1**-**3**, with all metal centers being hexacoordinated. It is interesting to note that the coordinating ligands are not arranged in the same manner around the $Mn_4O_6$ core as in the other derivative complexes. In **4**, the coordination sphere of each outer Mn(1) atoms is completed by one chelating protonated Hhmp ligand (See Figure 2). The protonation of these ligands is confirmed by the conformation of the oxygen atom O(4) that points to the intermolecular space and by bond-valence sum analysis. Moreover, the hydrogen atom linked to O(4) is involved in an hydrogen bond with a neighboring perchlorate anion with a O(4)···O(22A) bond distance of 2.779 Å. Although slightly more bent than in **1**-**3**, Jahn-Teller axis is present on the central Mn(2) atom. On basis of this observation and bond-valence sum analysis, Mn(1) and Mn(2) valences are respectively +2 and +3. In the structure, $[Mn_4]$ complexes are mutually related by a glide plane, thus inducing the presence of two orientations of the Jahn-Teller axis separated by an angle of 35.3°. The presence of two additional protonated Hhmp ligands leads to geometrical changes in the $Mn_4O_6$ core of **4** with respect to **1**-**3**. The main structural variations are found on the Mn-O-Mn angles and the Mn···Mn separations (See Table 2). For example, the central Mn(2)-O(1)-Mn(2) angle is about 3° larger than in the other complexes resulting in a Mn···Mn distance about 0.1 Å longer. Mn(1)–O distances are also slightly shorter in **4**, while only the Mn(1)-O(3)-Mn(2) angle is found larger in **4**. In the crystal packing, no significant intermolecular contacts (hydrogen bonds or π–π stacking) are found between adjacent [$Mn_4$] complexes.

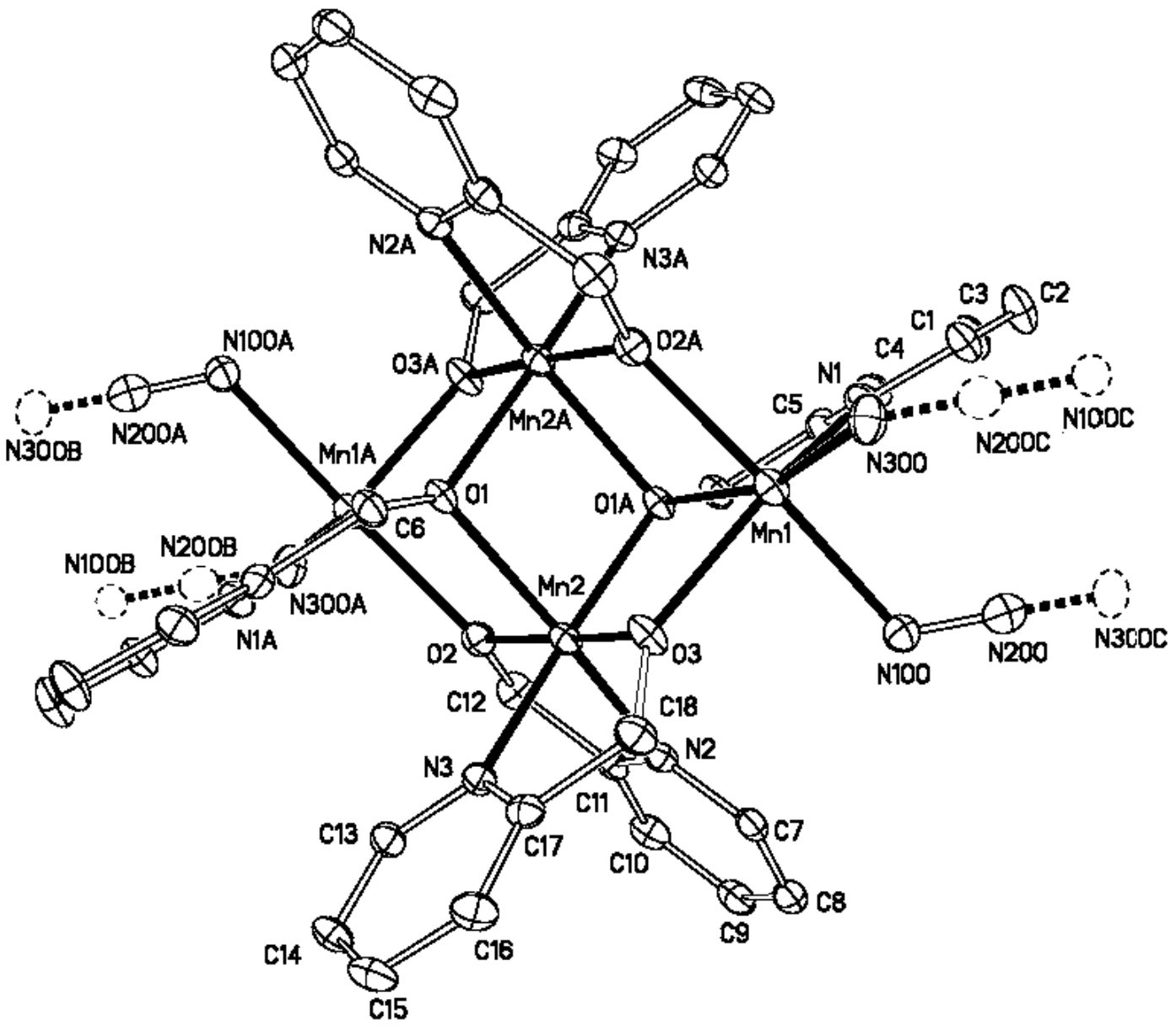

***Figure 3***. ORTEP representation of the repeating [$Mn_4$] unit in **5** with thermal ellipsoids at 30 % probability. Dashed bonds and atoms belong to neighboring [$Mn_4$] units. H atoms are omitted for clarity.

Compound **5**, catena-{[$Mn_4(hmp)_6(N_3)_2$]($ClO_4$)$_2$}, crystallizes in the monoclinic space group $P2_1/n$. Its structure is built from tetra-nuclear [$Mn_4(hmp)_6(N_3)_2$]$^{2+}$ units connected through two $N_3^-$ anions into a 1-D assembly. The tetra-nuclear core shown in Figure 3 is very similar to those in **1-4** with the outer Mn(1) atoms being hexa-coordinated. Table 2 compares bond angles and distances characteristic of the $Mn_4$ core for compounds **1**-**5**. As for the other compounds, the oxidation state of Mn(1) and Mn(2) is respectively assigned as divalent and trivalent based on bond distance and charge balance consideration. As expected, Jahn-Teller distortion is observed on Mn(2) sites. The [$Mn_4$] units are linked together through their $Mn^{II}$ ions by a double *end-to-end* (EE) azido bridge to form a linear arrangement along the *a* axis (see Figure 4). These double $N_3^-$ bridges have a chair-like conformation (Mn(1)-N-N angles are 118.7(5)° and 124.6(5)°, torsion between the two Mn(1)-N-N-N mean planes is 57.6(4)°, dihedral angle between the planes defined by the six N atoms of the azides and the N-Mn(1)-N system is 37.3(4)°). Within the chains, all [$Mn_4$] units, and thus Jahn-Teller elongation axes, have a unique orientation. On the other hand, two chain orientations are generated by symmetry with an angle of about 15°. Hence, this arrangement leads to the presence of two [$Mn_4$] orientations in the crystal structure. Although the chains are well separated by the counter-anions, very weak inter-chain contacts have been noticed through head-to-tail π-π interactions between hmp rings (c.a. 3.6-3.7 Å).

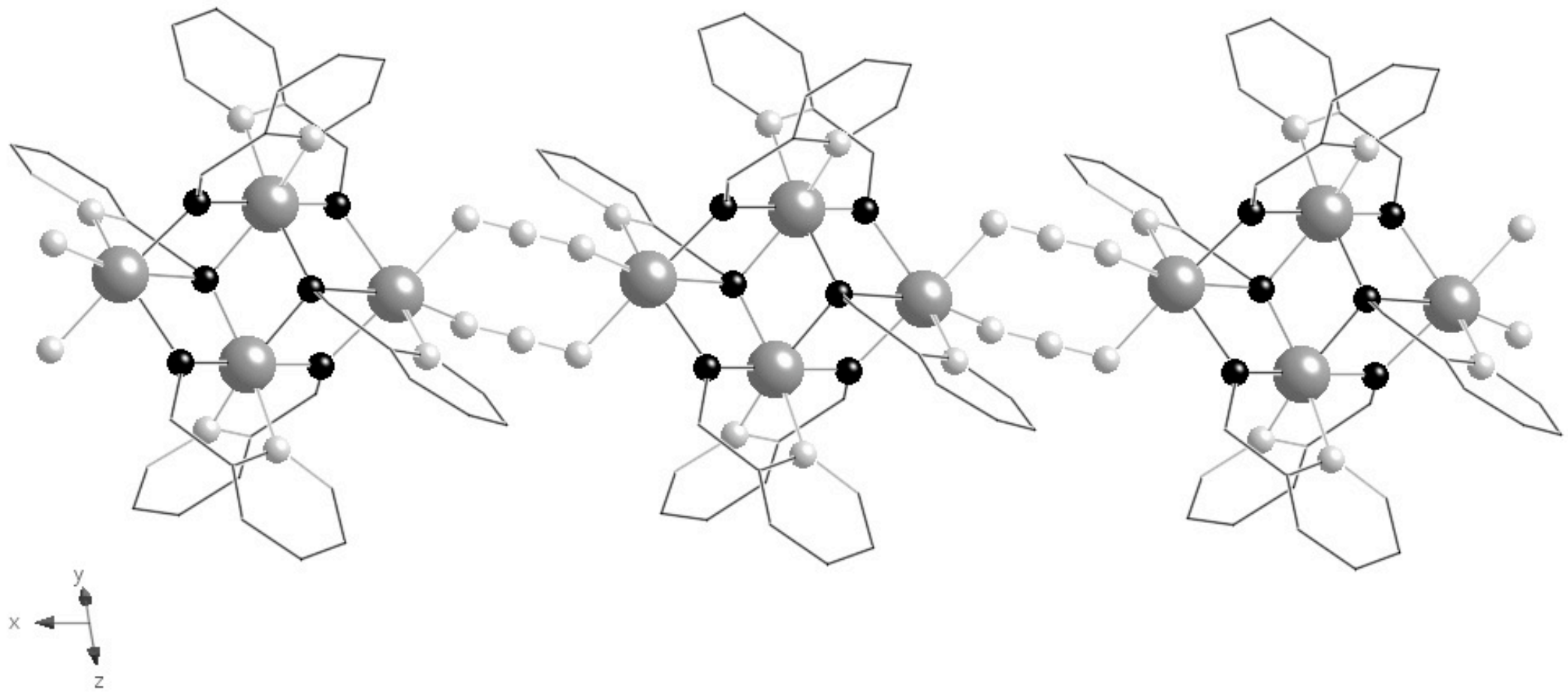

***Figure 4***. A view of the one-dimensional assembly of [$Mn_4$] units through $N_3^-$ bridges in the structure of **5**. Black, O; dark grey, Mn; light grey, N. H atoms and perchlorate anions are omitted for clarity.

**Table 2.** Comparison of the structural parameters of the [$Mn_4$] cores in compounds **1**-**5**.

| | | **1** | **2** | **3** | **4** | **5** |
|---|---|---|---|---|---|---|
| Jahn-Teller axis | Mn(2)–N | 2.1902(17) | 2.199(9) | 2.195(5) | 2.221(2) | 2.217(4) |
| | Mn(2)–O(1) | 2.2436(14) | 2.292(7) | 2.277(4) | 2.275(2) | 2.235(3) |
| | N-Mn(2)-O(1) | 160.10(6) | 163.8(4) | 159.24(15) | **151.24(9)** | 161.03(14) |
| Mn(2)-Mn(2) link | Mn(2)–O1 | 2.2436(14) | 2.292(7) | 2.277(4) | 2.275(2) | 2.235(3) |
| | | 1.9581(14) | 1.950(7) | 1.980(4) | 2.007(2) | 1.982(3) |
| | Mn(2)-O1-Mn(2) | 100.49(6) | 98.9(3) | 99.44(15) | **102.90(8)** | 100.17(14) |
| | Mn(2)···Mn(2) | 3.2354(10) | 3.230(3) | 3.253(2) | **3.353(2)** | 3.2389(16) |
| Mn(2)-Mn(1) links | Mn(1)–O(3) | 2.2472(14) | 2.148(7) | 2.201(4) | 2.149(2) | 2.144(4) |
| | Mn(1)–O(1) | 2.3712(14) | 2.290(8) | 2.275(3) | 2.247(2) | 2.239(4) |
| | Mn(1)–O(2) | 2.1761(14) | 2.179(8) | 2.185(4 | 2.141(2) | 2.197(3) |
| | Mn(2)–O(3) | 1.8601(14) | 1.890(7) | 1.881(3) | 1.881(2) | 1.859(4) |
| | Mn(2)–O(2) | 1.8749(14) | 1.875(8) | 1.860(3) | 1.873(2) | 1.875(4) |
| | Mn(2)···Mn(1) | 3.323(1) | 3.346(3) | 3.271(2) | 3.196(1) | 3.193(1) |
| | | 3.388(1) | 3.254(3) | 3.364(2) | 3.317(1) | 3.278(1) |
| | Mn(1)-O(1)-Mn(2) | 94.41(5) | 93.8(3) | 95.31(13) | 94.35(7) | 94.21(13) |
| | Mn(2)-O(2)-Mn(1) | 113.28(7) | 106.6(3) | 107.63(17) | 105.36(9) | 106.97(16) |
| | Mn(2)-O(1)-Mn(1) | 99.84(6) | 100.0(3) | 100.25(14) | 97.24(8) | 98.12(15) |
| | Mn(2)-O(3)-Mn(1) | 107.64(6) | 111.8(4) | 110.77(16) | 110.59(9) | 105.59(17) |

Data in bold highlight the main structural differences between **4** and the four other compounds of this study, at the origin of the different spin ground state.

### ***Magnetic properties of the tetra-nuclear complexes 1-4.***

Figure 5 shows the temperature dependence of the $\chi T$ product for compounds **2**-**4** (detailed data for **1** have been previously communicated[9]). Compounds **1**, **2** and **3** present a very similar behavior already observed for this type of tetra-nuclear complex:[8] $\chi T$ increases upon lowering the temperature from about 15 $cm^3mol^{-1}K$ at 300 K (in agreement with the expected value for uncoupled Mn(II) and Mn(III) ions i.e. 14.75 $cm^3mol^{-1}K$ for $g$ = 2), to a maximum of 40.5 / 46.2 / 38.7 $cm^3mol^{-1}K$ at 2.9 / 2.9 / 5.1 K respectively for **1**, **2** and **3**. At lower temperatures $\chi T$ then decreases down to 39.1 / 44.4 / 33.6 $cm^3mol^{-1}K$ at 1.83 K. This behavior is induced by ferromagnetic interactions among the Mn ions within the [$Mn_4$] clusters ($J_{bb}$ between $Mn^{3+}$ ions and, $J_{wb}$ between $Mn^{2+}$ and $Mn^{3+}$ ions), zero-field splitting (ZFS) of the high spin [$Mn_4$] cation and possible antiferromagnetic inter-complexes interactions. Experimental data can be reproduced using the Heisenberg-Van Vleck model already employed by G. Christou and D. N. Hendrickson *et al.* (the Hamiltonian used here is $\boldsymbol{H} = -2J_{bb}(\boldsymbol{S}_{Mn2}\boldsymbol{S}_{Mn2A}) - 2J_{wb}(\boldsymbol{S}_{Mn1}+\boldsymbol{S}_{Mn1A})(\boldsymbol{S}_{Mn2}+\boldsymbol{S}_{Mn2A})$).[8] Data below 13 K were omitted in the fitting procedure to avoid the ZFS effects or inter-complexes antiferromagnetic interactions. With temperature-independent paramagnetism (*TIP*) fixed at $6\times10^{-4}$ $cm^3mol^{-1}$, good fits were achieved, with final optimized parameters being $g$ = 1.96(1) / 2.01(1) / 1.93(1), $J_{wb}/k_B$ = +0.66(1) / +0.55(2) / +1.24(2) K, and $J_{bb}/k_B$ = +8.56(5) / +5.0(1) / +13.3(1) K for **1**, **2** and **3**, respectively. Clearly Mn ions in **1-3** are ferromagnetically coupled to give an $S_T$ = 9 ground state. The values of $J_{wb}$, $J_{bb}$ and $g$ are similar to those reported in literature.[8,9b]

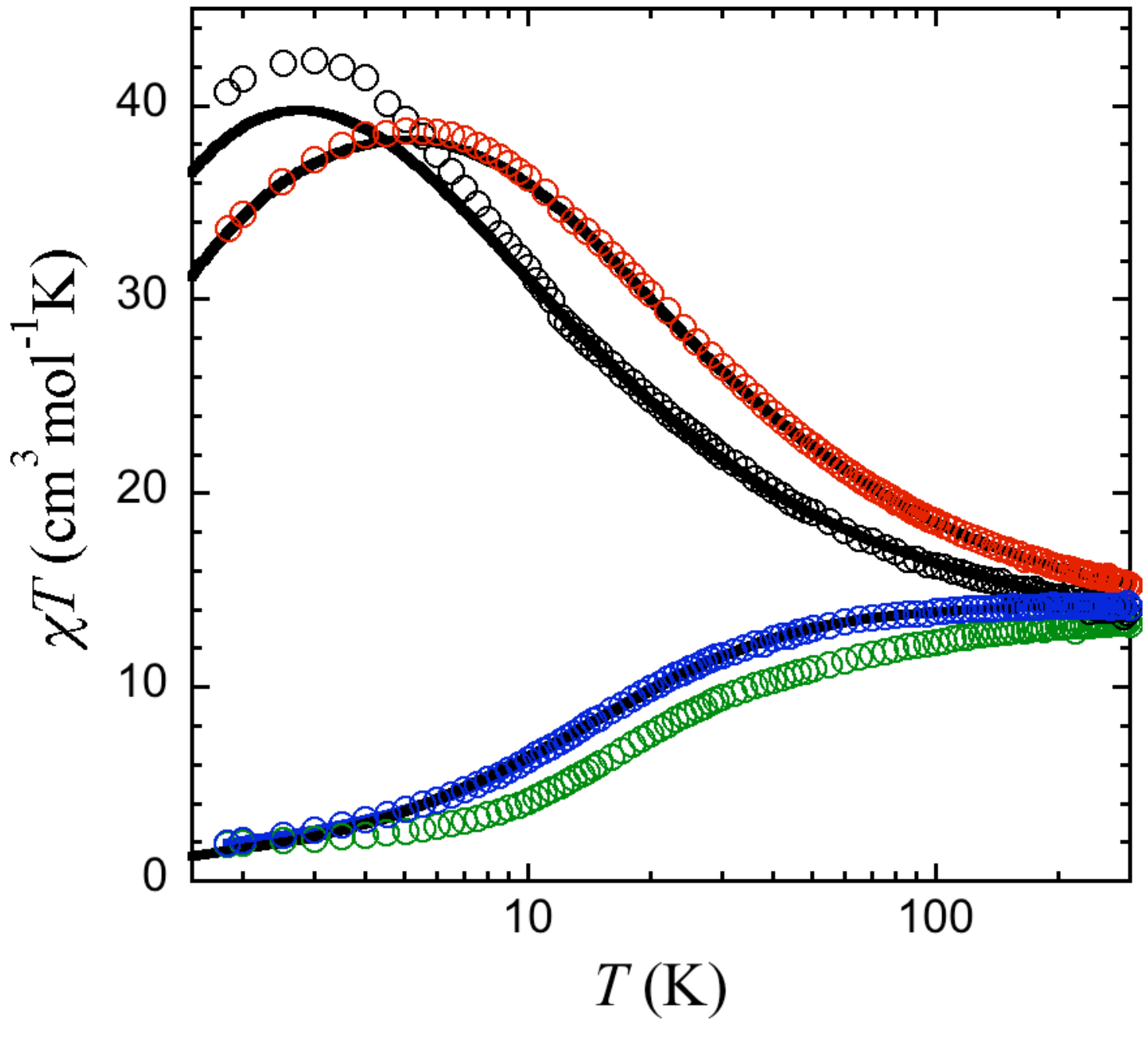

***Figure 5***. Plot of $\chi T$ *vs.* $T$ under 0.1 T for **2** (○), **3** (○), **4** (○) and **5** as-synthesized (○). The solid lines represent the best fit obtained with the tetranuclear models described in the text.

Isothermal field dependences of the magnetization have been measured in the average easy direction of **1** and **3** single crystals. A hysteretic response is observed with steps at regular magnetic field intervals (see Figure 6 for **3**). The coercive field varies strongly with temperature and becomes temperature-independent around 0.3-0.4 K indicating a purely *QTM* regime between $m_{ST} = \pm 9$. On the other hand, the regular steps correspond to resonant tunneling transitions between $m_S = 9$ and $m_S = -9$ (close to zero field), $m_S = -8$ (around 0.3 T), and $m_S = -7$ (around 0.6 T).

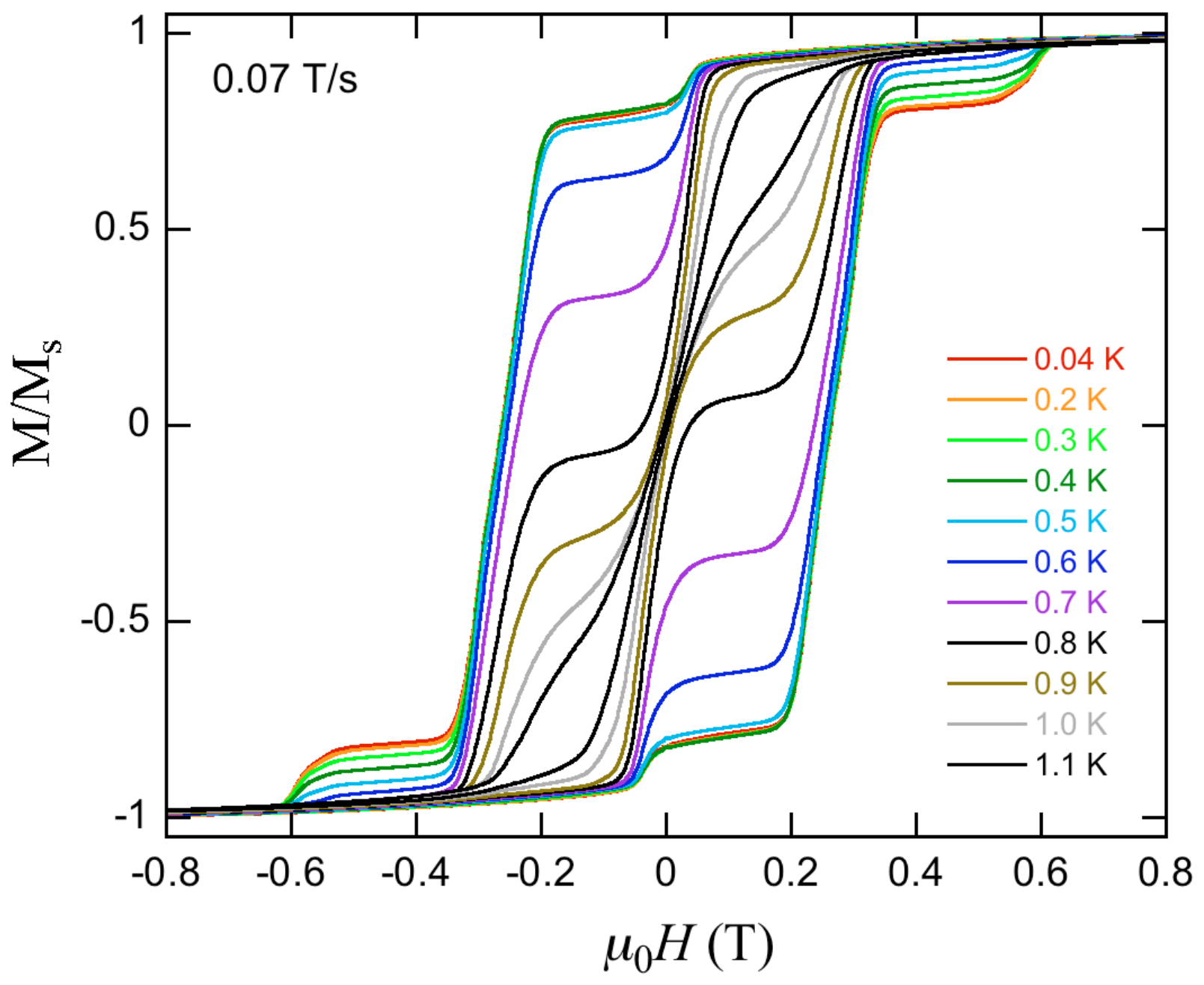

***Figure 6***. Magnetization vs. field hysteresis loops measured on **3** at different temperatures with a field sweep rate of 0.07 T/s. The data are normalized to the saturation magnetization ($M_s$) at 1.4 T.

As for previous [$Mn_4$] species,[8,9b] this feature is characteristic of the SMM behavior of **1** and **3**. The *D* parameter can thus be evaluated with a very good accuracy from the field separation between these steps ($\Delta H$), which is proportional to *D* such as $\Delta H^{(n)} = n|D|/(g\mu_B)$ (where *n* is an integer indicating the n$^{th}$ step). With $\Delta H = 0.25$ / 0.26 T and $g = 1.96$ / 1.93, $D/k_B$ is equal to –0.33 / –0.34 K respectively for **1** and **3**. The obtained *D* values are in good agreement with the previous estimation done in other derivatives.[8,9b] Given that the structural parameters at the origin of the $Mn^{III}$ anisotropy are very similar along the series of [$Mn_4$] complexes, a comparable value of *D* is expected in **2**. Based on the above *D* values, a numerical approach using Magpack program[17] has allowed us to estimate the local anisotropy $D_{Mn}/k_B$ of the $Mn^{III}$ sites at about –4.4 K. This value is in agreement with previous reports[9a,18] and is further confirmed by the reasonable simulation of the magnetic susceptibility at low temperatures displayed in Figure 5 ($J_{wb}$, $J_{bb}$ and *g* have been fixed to their value obtained from the high temperature fits and $D_{Mn}/k_B$ at –4.4 K). In the case of **3**, a small intermolecular interaction ($zJ'/k_B$ = –3 mK) had to be introduced in order to reach a correct simulation of the low temperature data.[19] A similar $zJ'/k_B$ value is obtained from the zero field shift of the first step in the magnetization hysteresis curve displayed Figure 7 ($H^*$ = –0.03 T, that corresponds to ca. –2 mK; $zJ' = g\mu_B H^*/(2S_T)$).

Regarding complex **4**, the $\chi T$ product decreases from 14.03 cm$^3$mol$^{-1}$K at 300 K down to 1.94 cm$^3$ K mol$^{-1}$ at 1.8 K, most of the drop occurring in the 80–2 K range. This original behavior within the $Mn_4$ family of compounds is indicative of significant intramolecular antiferromagnetic interactions. Indeed, using the same model as for **1**-**3**, the experimental data are well reproduced with the best-parameters $g$ = 2.05(1), $J_{wb}/k_B$ = –0.92(2) K and $J_{bb}/k_B$ = +0.25(5) K (Figure 5). The drastic decrease of $J_{bb}$ in **4** with respect to the three other $Mn_4$ complexes **1**-**3** is clearly related to wider Mn(2)-O(1)-Mn(2) angles and longer Mn(2)···Mn(2) separation in **4** (See Table 2). Such conditions favor orbital overlap and thus increase the antiferromagnetic component of this magnetic interaction. The change of sign of $J_{wb}$, at the origin of the dominantly antiferromagnetic behavior in **4**, corresponds to comparatively smaller geometrical variations. Slightly shorter Mn(1)–O distances with comparable Mn(1)-O-Mn(1) angles are found that probably favor a better orbital overlap. With these coupling scheme and parameters, the spin ground state in **4** is $S_T$ = 1 and no SMM behavior was observed.

As illustrated by the low temperature $M$ vs. $H$ hysteresis loops (Figure 6), the different [$Mn_4$] derivatives, **1**-**3**, display a SMM behavior and hence slow relaxation of their magnetization. The characteristic relaxation time was studied for **1**-**3** using ac measurements above 1.8 K (Figure S2-S4) and dc $M$ vs. $t$ measurements below 1.80 K (for **1** and **3**, Figure S5-S6). The relaxation time $\tau$ was extracted from the maximum of $\chi$'' (either vs. $T$ or vs. $\nu$) in ac measurements and taking $\tau = t$ when $M/M_s(t)$ reaches $1/e$ in dc measurements. In the ac temperature domain, the relaxation time ($\tau$) is thermally activated with $\Delta_{eff}/k_B$ = 23.3 / 23.2 / 19.6 K and $\tau_0$ = 9.7 / 12 / 4.4 × 10$^{-10}$ s (inset Figure 7) for **1**, **2** and **3**, respectively. Experimentally, in this temperature domain, two relaxation processes (thermal and quantum) are in competition. Hence, the thermal barrier is slightly "short-cut" by the quantum tunneling of the magnetization and the effective energy barrier ($\Delta_{eff}$) on the relaxation time is smaller than the theoretical one. In **1** and **2**, $\Delta_{eff}$ is surprisingly high for a compound of this family (usually found around 15-18 K),[8a-b] but still lower than the expected barrier: $\Delta/k_B = |D|S_T^2/k_B = 26.7$ K. Below 1 K, the relaxation time does not follow anymore an Arrhénius law and saturates slowly as expected when quantum tunneling of the magnetization (*QTM)* becomes the fastest pathway of relaxation. As already suggested by *M* vs. *H* data (Figure 6) and shown on Figure 7, a *QTM* regime is observed below 0.35 / 0.43 K with a characteristic time $\tau_{QTM}$ = 6900 / 400 s for **1** and **3** respectively. As recently established for $[Mn_4(hmp)_6(H_2O)_2(NO_3)_2](NO_3)_2$•$2.5H_2O$ using Landau-Zener method,[9b] the high $\Delta_{eff}$ and $\tau_{QTM}$ values for **1** contrasts with the 15.8 K and 1000 s measured for $[Mn_4(hmp)_6Br_2(H_2O)_2]Br_2$•$4H_2O$[8c] suggesting that the transverse anisotropy (*E*), which governs $\tau_{QTM}$, is significantly reduced in **1**, although the $Mn_4O_6$ cores possess the same local $C_{2v}$ symmetry. Extrapolating this comment to other $Mn_4$ complexes described in this paper, **1** and **2** would have an *E* value comparable to the one observed in $[Mn_4(hmp)_6(H_2O)_2(NO_3)_2](NO_3)_2$•$2.5H_2O$ ($E/k_B$= +0.083 K)[9b] but lower than in **3** and $[Mn_4(hmp)_6Br_2(H_2O)_2]Br_2$•$4H_2O$ ($E/k_B$ = +0.124 K)[8c].

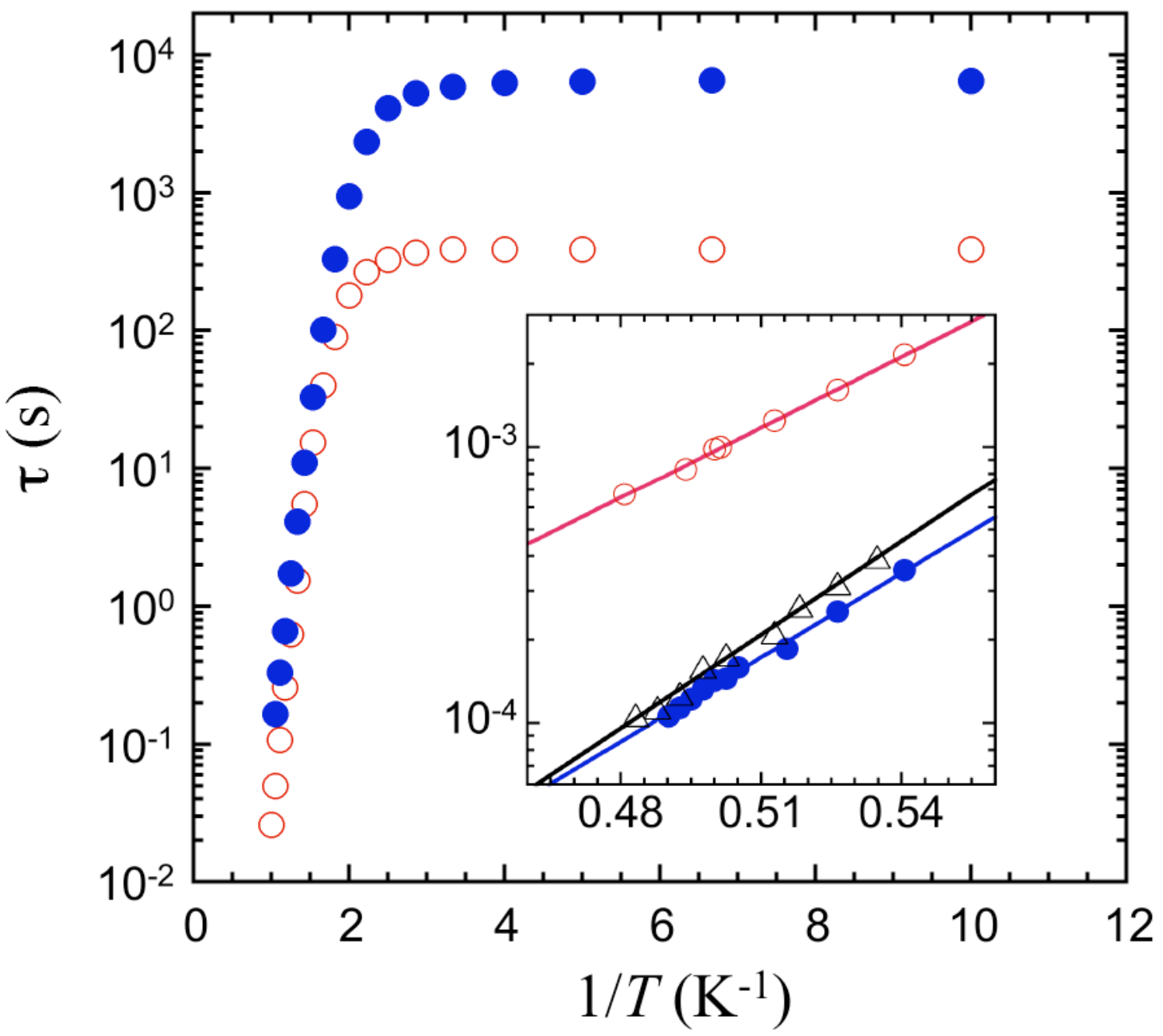

***Figure** 7*. Semi-log plot of the relaxation time $\tau$ vs. $1/T$ determined by dc measurements on **1** (●) and **3**(○). Inset: $\tau$ vs. $1/T$ plot determined from ac susceptibility data for **1** (●), **2** (Δ) and **3**(○). The solid lines represent Arrhénius fits of the high temperature data.

### *Magnetic properties of the chain compound (5)*

**Static properties.** The $\chi T$ product for **5** decreases first slightly from 13.22 $cm^3$ K $mol^{-1}$ at 300 K down to 12.24 $cm^3$ K $mol^{-1}$ at 100 K and then drops in a more pronounced manner in the range 60-10 K down to 2.4 $cm^3$ K $mol^{-1}$ at 2 K (Figure 5). This thermal variation is completely different than those of the isolated [$Mn_4$] complexes **1**-**3** but is reminiscent of the magnetic behavior observed for **4**. As indicated by the decrease of $\chi T$ product with temperature, dominating antiferromagnetic interactions are present in **5**. Nevertheless, an in-depth comparison analysis of the [$Mn_4$] core structural parameters in **1**-**4** and **5** (see Table 2) is not in favor of intra-[$Mn_4$] antiferromagnetic interactions. On the other hand, magneto-structural data for Mn(II)-azido systems confirms that a weak antiferromagnetic interaction is expected for $Mn^{II}$-$(N_3)_2$-$Mn^{II}$ bridges such as those found in **5**.[20] We have thus used the susceptibility expression for [$Mn_4$] units[8a] and considered the interaction through the azido bridges as an inter-tetranuclear interaction $J'$, treated in the mean-field approximation.[19] Considering the structural similarities between the [$Mn_4$] core of **1** and **5**, we have fixed $J_{bb}$ and $J_{wb}$ values to those obtained for **1** ($J_{bb}/k_B$ = +8.6 K, $J_{wb}/k_B$ = +0.7 K). Down to 15 K, the data are indeed correctly reproduced (Figure S7) within these conditions with a small negative $zJ'$ (ca. $J'/k_B \approx -0.15$ K with $z$ = 2).[19,21] Interestingly, this intermolecular interaction would correspond to an Mn–Mn interaction of ca. –1.9 K,[22] value falling in the range of couplings observed between $Mn^{II}$ ions bridged through end-to-end $N_3^-$.[20] Therefore, **5** can be viewed a one-dimensional arrangement of antiferromagnetically coupled [$Mn_4$] units. As shown on the complexes **1**-**3**, these high-spin moieties possess a strong uni-axial anisotropy that should induce at low temperature a magnetic behavior for **5** compatible with a chain of antiferromagnetically coupled Ising $S_T$ = 9 spins. As expected from this chain model, the ln($\chi T$) vs. 1/$T$ plot decreases almost linearly between 20 and 10 K confirming the one-dimensional nature of **5**, the presence of antiferromagnetic interactions and an uni-axial anistropy (Figure 8). At lower temperature, ln($\chi T$) does not really saturate and a residual component is still observed. This low temperature behavior is systematically observed in the different synthesized samples but its intensity varies and is minimized after recrystallization of the material (see Figure 8).[23] A residual component is actually expected for a chain of antiferromagnetically coupled spins as the perpendicular component of the susceptibility ($\chi_\perp$) becomes dominant at low temperature. Nevertheless, if this term has to be considered, it does not explain the sample-dependent behavior that likely arises from an additional paramagnetic contribution dependent on the compound crystallinity. This point is further confirmed by the $\chi T$ vs. $T$ plot, displayed as inset in Figure 8, that shows $\chi T$ products which do not extrapolate to zero at 0 K as expected. From the 0 K extrapolation, a lower-limit estimation of Curie site percentage can be done in the as-synthesized compound at about 4.5 % and after recrystallization at about 3.5 % (considering that these paramagnetic sites possess the Ising Curie constant of an $S_T = 9$ spin: $C = 40.5$ cm$^3$ K mol$^{-1}$). Note that by considering the $\chi T$ value at 1.8 K, an upper-limit estimation of Curie site percentage is obtained at 6.5 and 5.0 % respectively.

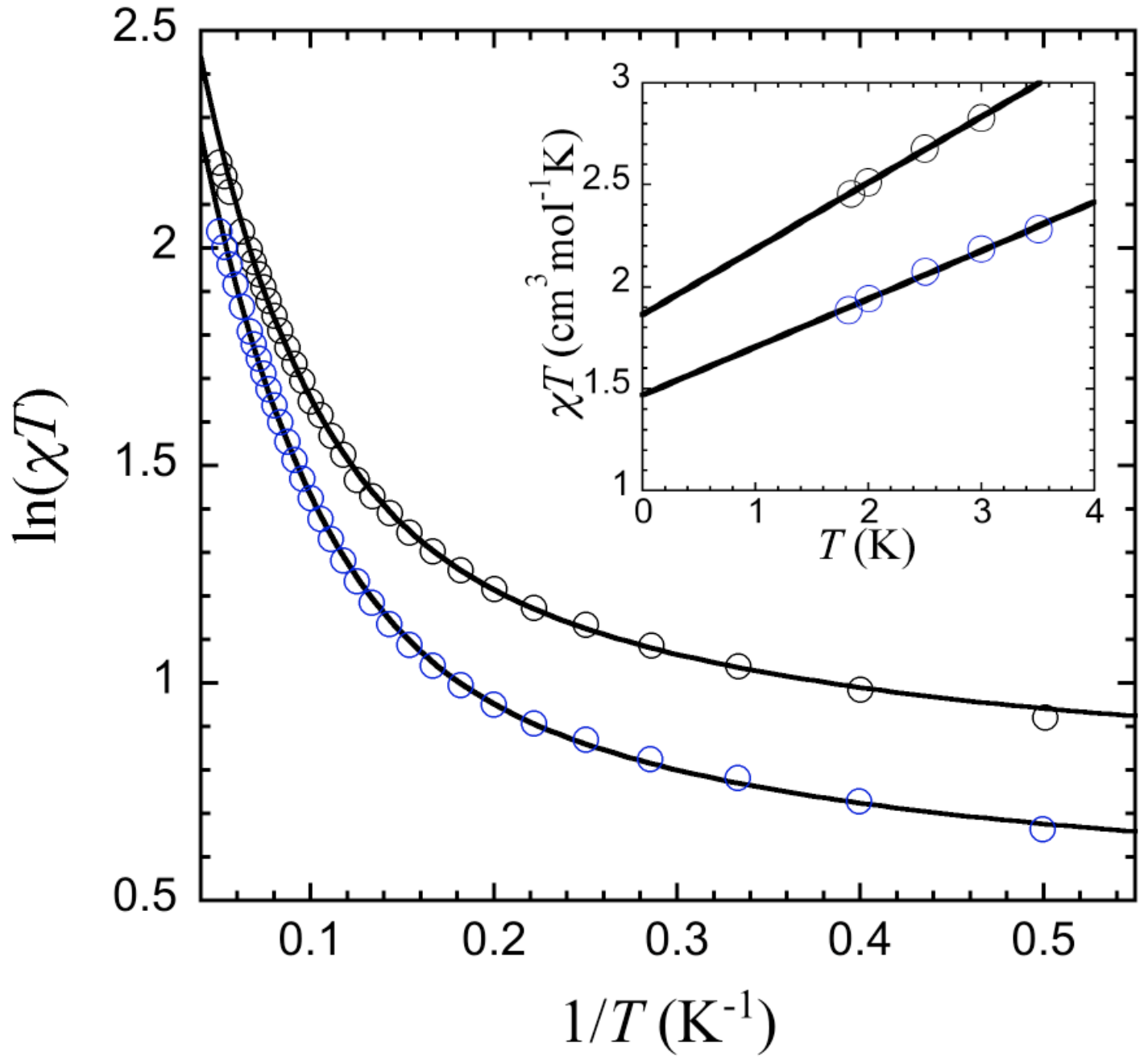

***Figure 8***. Plot of ln($\chi T$) vs. $1/T$ for **5** in the range 20-2 K for as-synthesized (○) and recrystallized (○) samples. Full lines represent the fit to a model of finite-size chains of antiferromagnetically-coupled Ising spins (see text). Inset: $\chi T$ vs. $T$ plot of the data from dc measurements for both samples.

This paramagnetic response has been predicted when finite-size effects are relevant in chains of antiferromagnetically coupled Ising spins.[24] This contribution is explained by the non-compensated magnetization resulting from segments composed of odd number of spins that possess a magnetization equivalent to one magnetic site (here $S_T = 9$). Experimentally, these defects can result from missing links between magnetic [$Mn_4$] complexes, i.e. a missing or modified double azido bridge. The percentage of missing links is thus equal to the number of segment i.e. twice the number of odd segments (from 13 to 9 and from 10 to 7 % from the above estimations) considering an equal probability to have a segment with an odd or even number of spin. Using the work of F. Matsubara et al, we have fitted the experimental susceptibility below 20 K considering the following expression:

$$\chi T = \chi_{//}/3 + 2\chi_{\perp}/3 = C\Big((1-x)\exp\big(4J'S_T^2/k_B T\big) + x/2\Big)\Big/3 + 2\chi_{\perp}T/3$$

where $\chi_{//}$ and $\chi_{\perp}$ are the parallel and perpendicular susceptibility, $x$ is the concentration of missing links and $C$ is the Curie constant for a magnetic site.[25] If all these parameters are let free, the fitting procedure does not converge. Therefore, the Curie constant $C$ was fixed at 32 $cm^3mol^{-1}$, the experimental $\chi T$ value taken as an average Curie constant for an isolated [$Mn_4$] unit in the considered temperature range (see Figure 5). With a fixed $C$ value, this model reproduces well the experimental data as seen on Figure 8. $\chi_{\perp}$ varies from 0.35 $cm^3mol^{-1}$ for the as-synthesized sample to 0.26 $cm^3mol^{-1}$ for the recrystallized sample.[26] We would have expected to find in both cases the same $\chi_{\perp}$ value but the magnetization slow relaxation at low temperature (vide infra) makes the fit inaccurate below $T$ = 3 K when this parameter is dominating the magnetic behavior. The amount of missing links, $x$, arises to 33 and 27 % respectively for the as-synthesized and recrystallized samples. These values seem highly overestimated in comparison to those deduced from the $\chi T$ vs. $T$ plot (vide supra). The estimation of the gap, $4J'S_T^2/k_B$, seems less ambiguous as a single value of –26(1) K is found for both as-synthesized and recrystallized samples. The deduced inter-complexes interaction, $J'/k_B \approx -0.08$ K, is in very good agreement with the $J'$ value obtained with the high temperature fitting of the magnetic susceptibility. Nevertheless, the whole set of parameters ($J'$, $\chi_{\perp}$ and $x$) will be further commented based on the low temperature dynamic behavior.

**Low temperature dynamic behavior: observation of a magnetization slow relaxation.** Low temperature dc μ-SQUID measurements have been performed on single crystals of compound **5** down to 1.3 K. As seen on Figure 9, field hysteresis loops of the magnetization appear below 2 K with a coercive field that is both field-sweep rate and temperature dependent. This type of hysteretic behavior is associated with the presence of magnetization slow relaxation. It is important to note that the shape of these hysteresis loops is drastically different from the one displayed in Figure 6 for [$Mn_4$] SMM complexes. This result strongly suggests that the present behavior is not coming from isolated SMMs and that the magnetic correlations along the chain participate actively to this slow relaxation as observed in Single-Chain Magnets.[5-7,27]

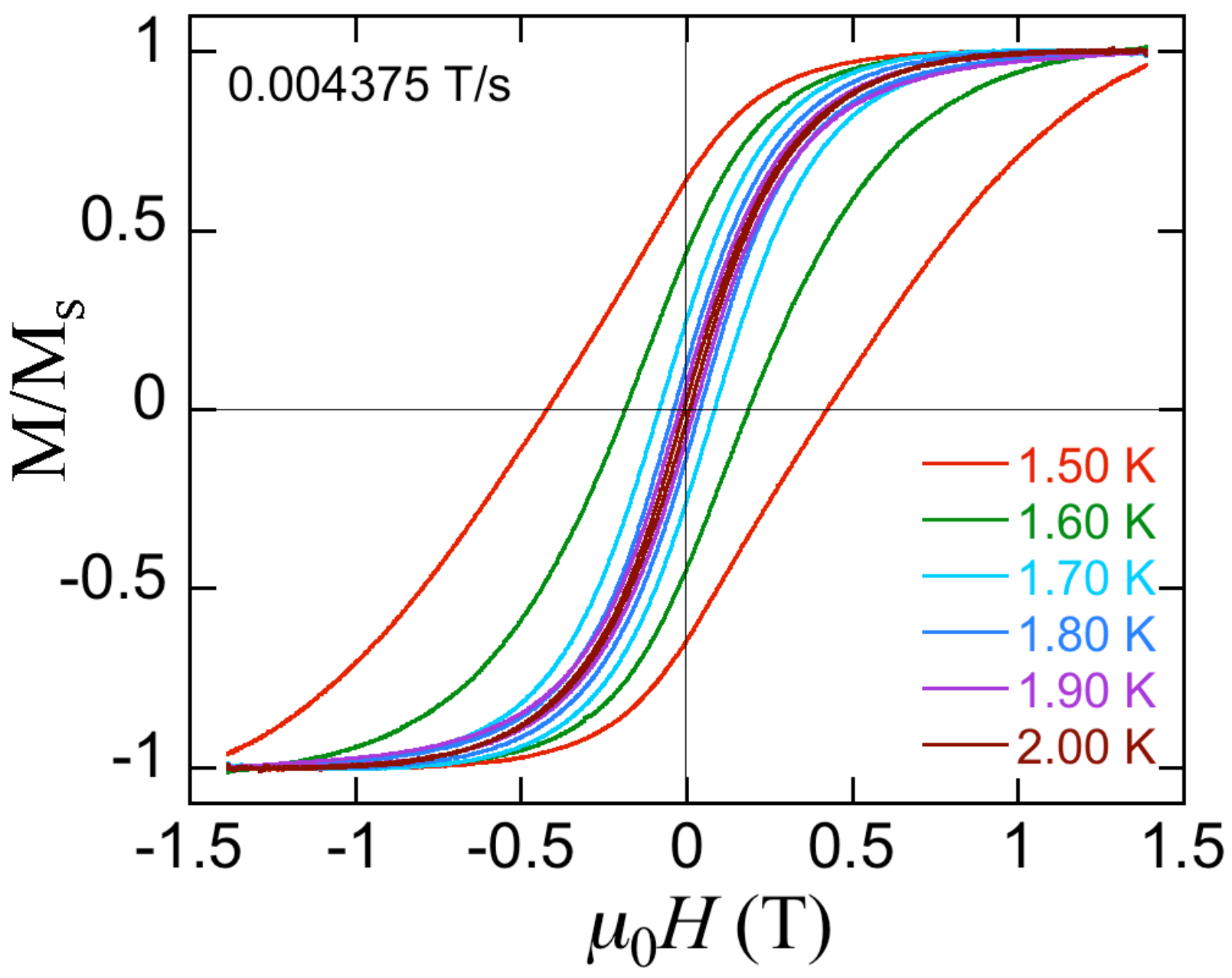

***Figure 9***. Magnetization vs. field hysteresis loops for **5** with a field sweep rate of 0.004375 T/s at various temperatures.

To probe this dynamics, ac susceptibilities were thus also measured on compound **5**. As shown in Figure 10, the ac susceptibility is frequency-dependent with a single relaxation mode. As expected due to the perpendicular contribution, an offset on the in-phase susceptibility ($\chi$') is observed even at the highest frequency available.[28] The amplitude of the relaxation deduced from both component of the ac susceptibility allows an estimation of $x$ of about 18 % and 8 % in the as-synthesized and recrystallized samples, respectively.[29] These $x$ values are in relatively good agreement with the estimation made from the $\chi T$ vs. $T$ data and confirm the over-estimation made from the fit of the ln($\chi T$) vs. 1/$T$ plot. On the basis of the magnetic measurements, these defects (missing links) are always present in the different batches of compound **5**. Interestingly, their amount is divided by about a factor 2 upon recrystallization.[23] The presence of these defects could explain the crystallization problems mentioned in the synthetic part and thus the limited size of the single-crystal that we were able to obtain.

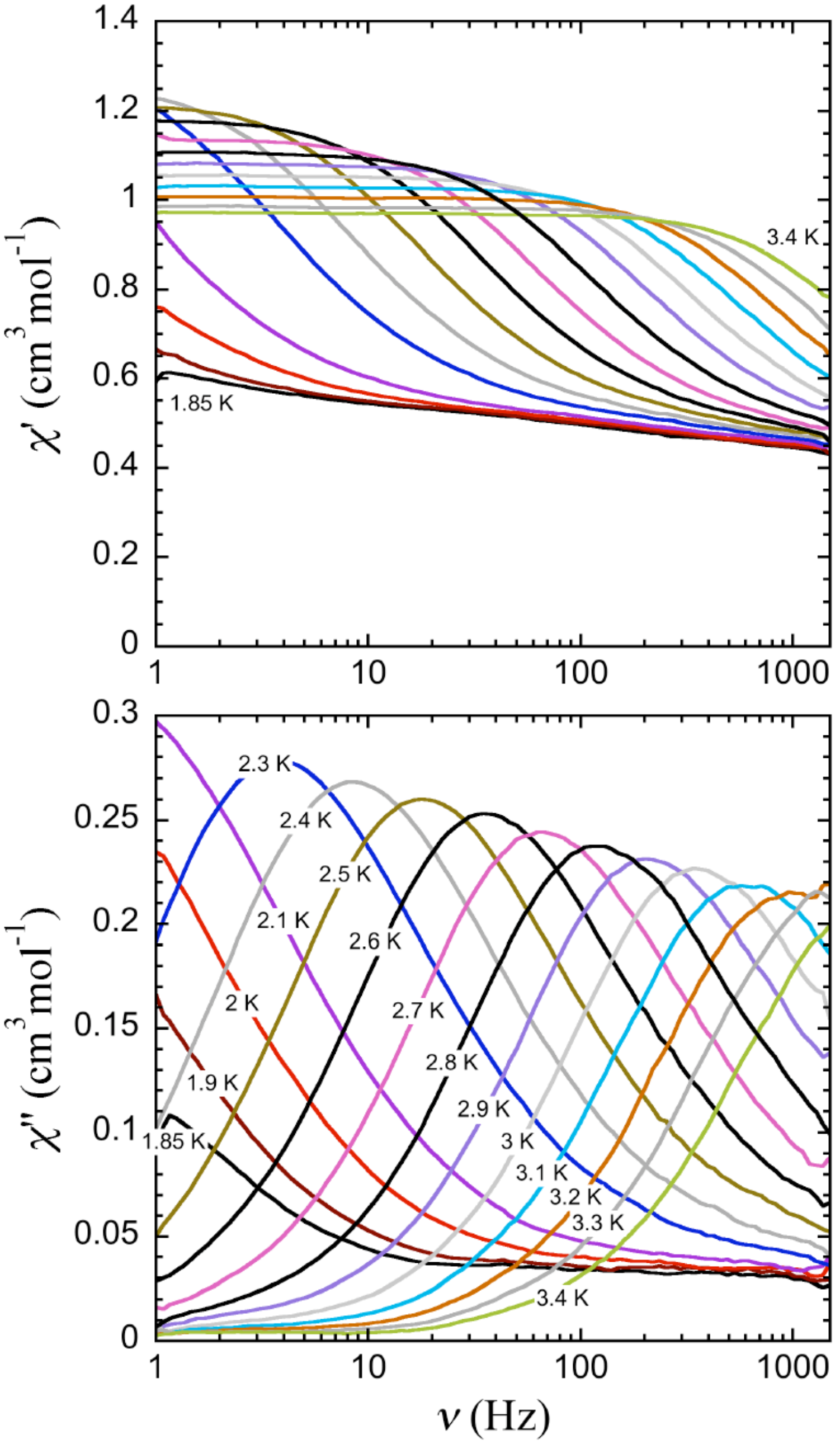

***Figure 10.*** Plot of in-phase ($\chi'$) and out-of-phase ($\chi''$) component ac susceptibility as a function of the frequency ($\nu$) for **5** (as-synthesized sample) under zero dc field.

Lowering the temperature, the relaxation time becomes too slow to be studied with ac measurements in the range of available frequencies. Therefore direct measurements of the magnetization relaxation were performed down to 1.3 K (Figure 11) showing the presence of a single relaxation mode. The relaxation time $\tau$ was extracted at each temperature taking $\tau = t$ when $M/M_S(t)$ reaches 1/e.

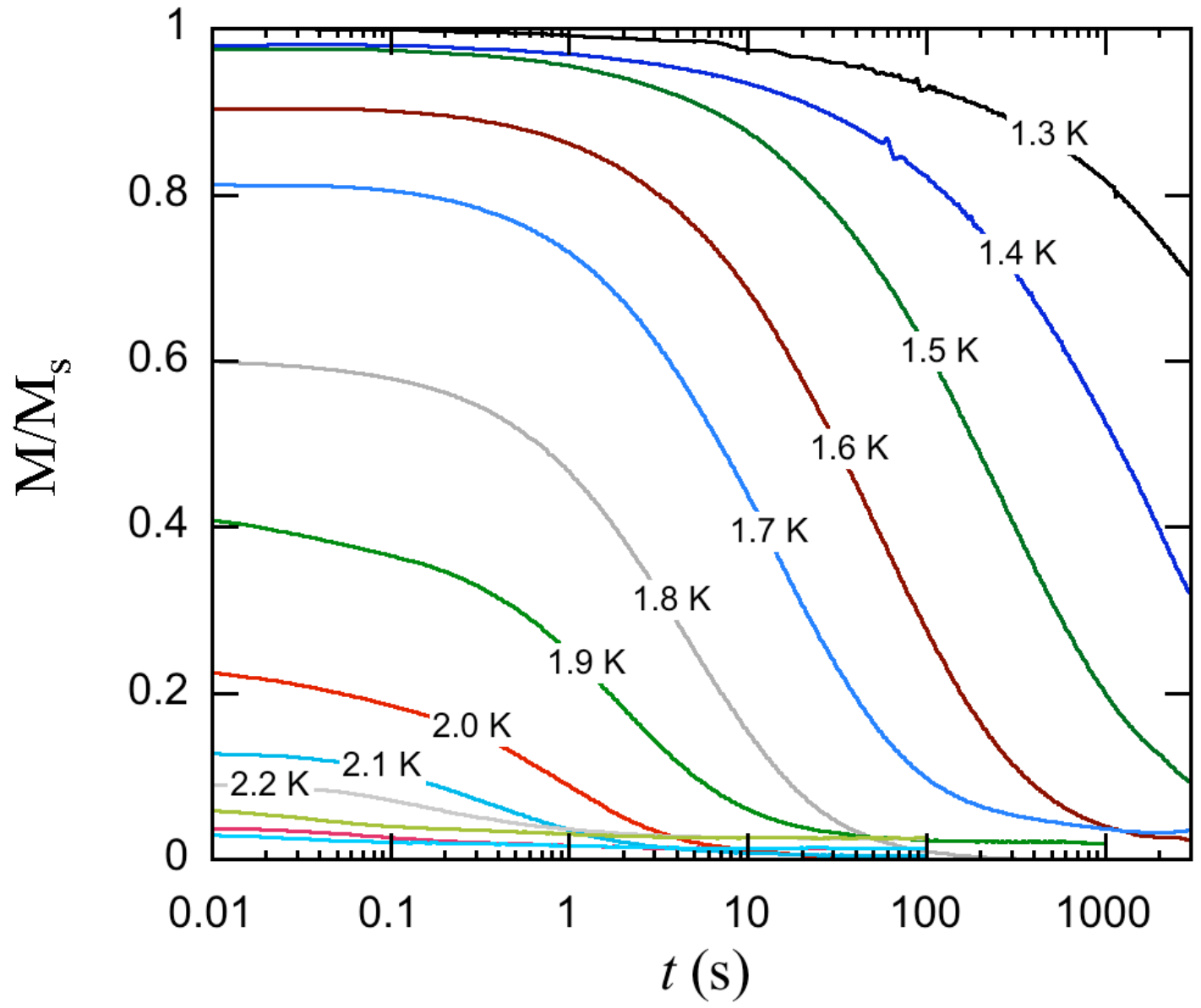

***Figure 11***. Relaxation of the magnetization of **5** at different temperatures. The data are normalized to the saturation magnetization ($M_s$) at 1.4 T.

In the resulting Arrhenius plot comparing data for **5** and **1** (Figure 12), it is evident that the range of relaxation time observed in **5** is completely different than in [$Mn_4$] SMMs. Remarkably, for an identical temperature, the relaxation time is 3 to 4 orders of magnitude higher in the chain compound **5** than in **1**. In addition, the low temperature dc data follow exactly the same Arrhenius law as the high temperature ac data with $\Delta_\tau/k_B$ = 47 K and $\tau_0 = 7\times10^{-11}$ s. Therefore no indication of quantum tunneling has been observed down to 1.2 K and below $10^5$ s although pure quantum tunneling of the magnetization is usually observed for SMM [$Mn_4$] species above $10^4$ s.[8c,9] Moreover, the observed energy gap of 47 K is about twice larger in **5** than in the SMM complexes. The origin of the magnetization relaxation observed in **5** can thus only be coming from the chains.

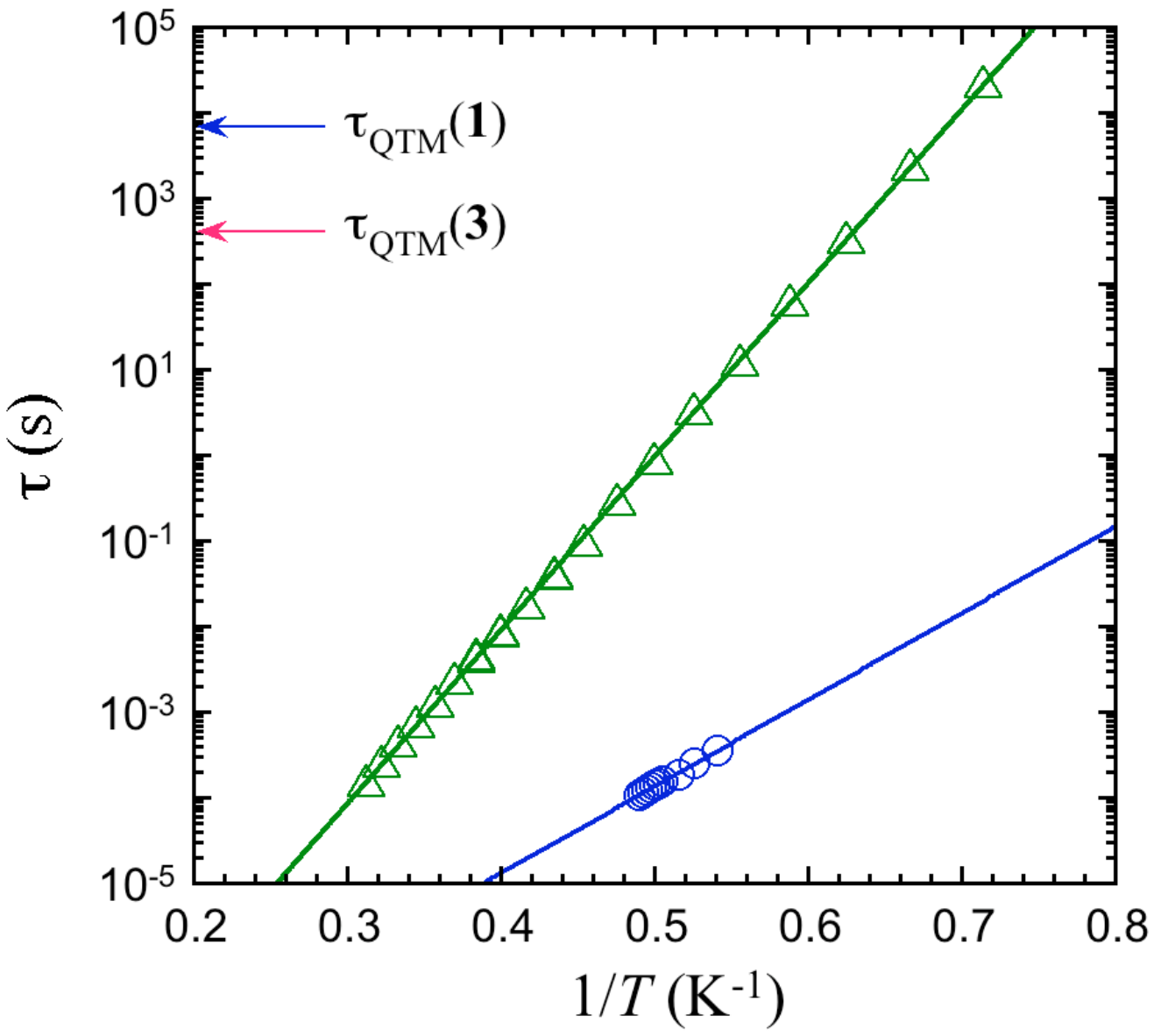

***Figure 12***. Semi-log plot of the relaxation time $\tau$ vs. $1/T$ determined by dc and ac techniques for **5** (△). The solid line represent an Arrhénius fit of all data. Data for **1** from ac susceptibility (○) with the corresponding Arrhenius fit are also presented for comparison. Arrows indicate the $\tau_{QTM}$ for the tetranuclear SMMs **1** and **3**.

The dynamics of antiferromagnetic chains have been described by M. Suzuki and R. Kubo.[30] In this case, the slow relaxation concerns the staggered magnetization, i.e. a mode corresponding to the alternation of the neighboring spins orientation. Therefore no neat magnetization can exist for an infinite antiferromagnetic chain. However the problem is qualitatively different if defects are introduced to cut the chain into finite segments containing either an odd or even spin number. In the former case, the magnetization of a segment is equal to the magnetic moment of an individual spin unit. Although, this kind of segment experiences a slow relaxation of the staggered magnetization, this dynamics can be probe by the total magnetization of this segment. The corresponding relaxation time has been calculated by J. H. Luscombe et al for a chain segment of a given size.[31] When defects are randomly distributed along the chain, a distribution of segment sizes should however be considered. The corresponding poly-disperse model has been discussed in the ferromagnetic case by D. Dhar and M. Barma.[32] Using the steepest descent approximation, they found an original time dependence for the relaxation of the magnetization. Except at very short or long times, they predict that the magnetization should decay as $\exp(-\sqrt{t/\tau})$ which contrasts with the simple exponential dependence obtained when a single segment size is considered. This result is also true in the antiferromagnetic case.[33]

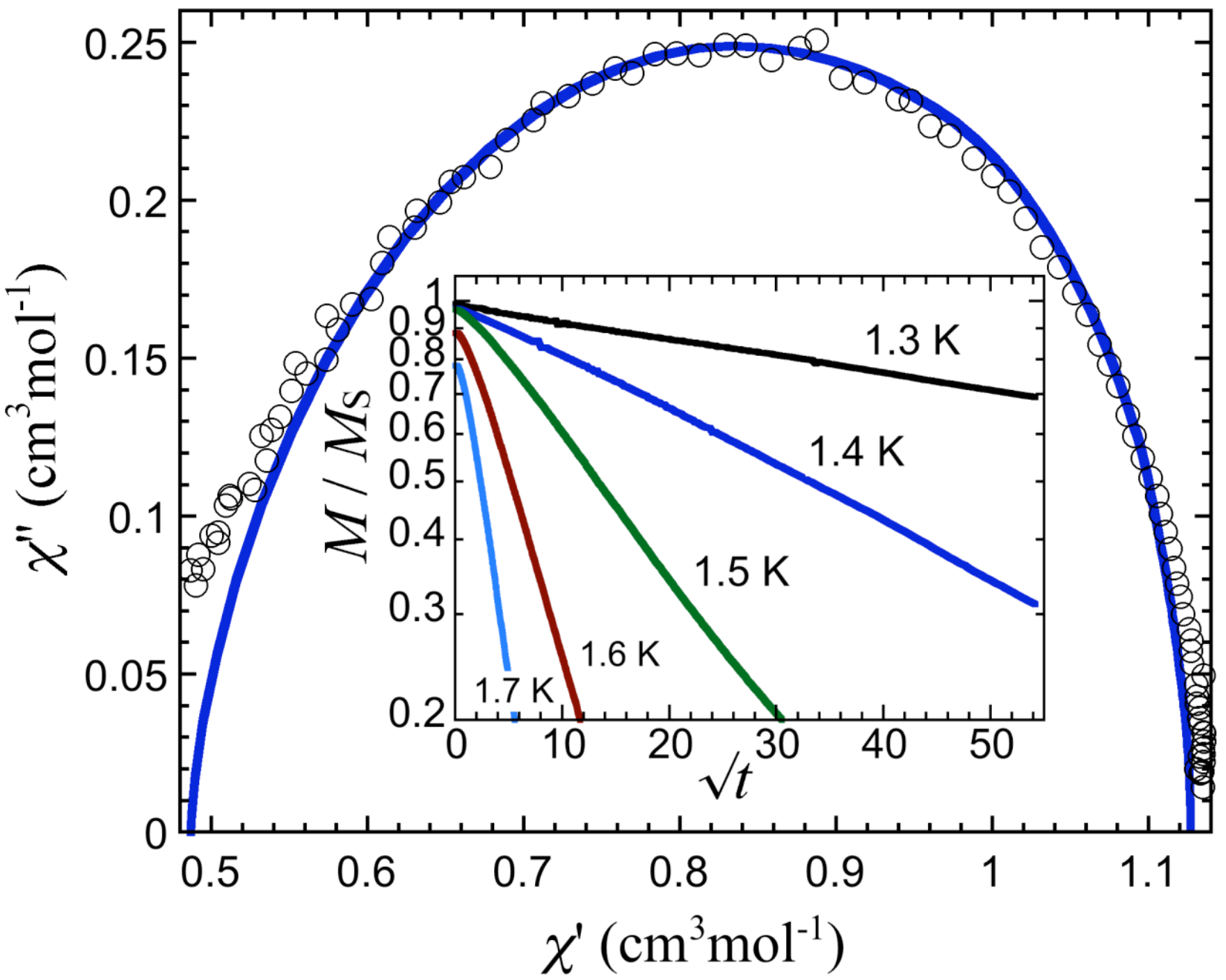

***Figure 13***. Cole-Cole plot deduced from the ac susceptibility components at 2.7 K (○) (data shown in Figure 10). The solid line gives the theoretical prediction from the poly-disperse model. To compare with experiment, the normalized theoretical susceptibility components have been rescaled as: $\chi' = a\chi'_{nor} + b$ and $\chi'' = a\chi''_{nor}$, with a = 0,64 cm$^3$ mol$^{-1}$ and b = 0,48 cm$^3$ mol$^{-1}$. Selected data from Figure 11 are plotted in inset as a function of $\sqrt{t}$.

The inset of Figure 13 shows that experimental relaxation data for **5** support the prediction of the poly-disperse model, i.e. straight lines are obtained when the logarithm of the magnetization is plotted as a function of $\sqrt{t}$. The measured ac susceptibility can also be compared with the theoretical ac susceptibility that we have calculated numerically.[34] Comparing with the well-known Debye model, the most striking difference is in the Cole-Cole plot that displays an unsymmetrical shape. As shown in Figure 13, this asymmetry is indeed manifest in the experimental results that are correctly reproduced. For clarity, we have only shown the experimental data obtained at 2.7 K but a similar agreement is found for other temperatures. Therefore, both ac data and direct measurements of the magnetization relaxation support a poly-disperse description, as expected when defects are randomly distributed along the chain. As far as we know, this is the first experimental confirmation for the $\sqrt{t}$ dependence of the magnetization relaxation predicted by D. Dhar and M. Barma.[32] Let us now discuss the temperature dependence of the relaxation time. The mono-disperse and poly-disperse models predict the same relaxation time gap coming from the magnetic correlations, i.e. $\Delta_\xi = 4|J'|S_T^2$. Considering the contribution from the individual relaxation of an isolated SMM unit ($\Delta_A$), the total gap for the relaxation time should be $\Delta_\tau = \Delta_A + \Delta_\xi = (4|J'|+|D|)S_T^2$. Taking into account the experimental energy gap for a $[Mn_4]$ SMM unit (typically $\Delta_A/k_B \approx 23$ K for **1** and **2)** and $\Delta_\xi/k_B \approx 26$ K obtained from the fitting of $\ln(\chi T)$ vs. $1/T$ data, the expected $\Delta_\tau = \Delta_A + \Delta_\xi$ reaches 49 K, in excellent agreement with the observed experimental value $\Delta_\tau/k_B = 47$ K.

The observation of magnetic slow relaxation in an antiferromagnetic Ising-like chain is an unprecedented result that highlights the paramount role of the finite-size effects (or defects) to probe this type of dynamics. In other words, it is only because this material possesses intrinsic defects (probably missing double azido bridges) that this behavior has been detectable.

## Concluding Remarks

We have shown in this paper synthetic and structural evidence that external coordination positions of $Mn_4$/hmp rhombic SMMs are easily exchangeable, and that using bridging bidentate coordinating anion such as the azido anion, an uni-dimensionnal assembly of the $[Mn_4]$ SMM can be obtained in a controlled manner. The magnetic properties of three $[Mn_4]$ complexes, $[Mn_4(hmp)_6(CH_3CN)_2(H_2O)_4](ClO_4)_4{\cdot}2CH_3CN$ (**1**) $[Mn_4(hmp)_6(H_2O)_4](ClO_4)_4{\cdot}2H_2O$ (**2**) and $[Mn_4(hmp)_6(H_2O)_2(NO_3)_2](ClO_4)_2{\cdot}4H_2O$ (**3**), have been studied, revealing their SMM behavior induced by an $S_T = 9$ ground spin and strong uniaxial anisotropy. On the other hand small structural variations resulting from the occupation of terminal positions by protonated Hhmp ligands in $[Mn_4(hmp)_6(Hhmp)_2](ClO_4)_4{\cdot}2CH_3CN$ (**4**) results in an $S_T = 1$ ground state. On basis of various dc and ac magnetic measurements, the chain compound catena-$\{[Mn_4(hmp)_6(N_3)_2](ClO_4)_2\}$ (**5**) can be viewed as a chain of antiferromagnetically coupled $S_T = 9$ anisotropic spins with an intra-chain interaction $J'/k_B$

≈ –0.1 K. The presence of finite-size chains in **5**, as a result of structural defects, allows the observation of slow relaxation of magnetization with an activated relaxation time ($\Delta_\tau/k_B$ = 47 K and $\tau_0$ = 7×10$^{-11}$ s). The experimental energy gap of $\tau$ has been successfully rationalized based on the individual slow relaxation of the $[Mn_4]$ unit (SMM) and the magnetic one-dimensional correlations of the system. Such observation is to date unique and demonstrates that the slow relaxation of antiferromagnetic Ising-like chains can be studied using intrinsic defects.

***Acknowledgements.*** This work was supported by PRESTO, CREST, Grant-in-Aid from the Ministry of Education, Culture, Sports, Science, and Technology (Japan), the EC-TMR Network "QuEMolNa" (MRTN-CT-2003-504880), the CNRS, the University of Bordeaux 1 and the Conseil Régional d'Aquitaine.

***Supporting Information Available:*** Additional magnetic data for the reported compounds together with X-ray crystallographic file of **2-5** in CIF format. This material is available free of charge via Internet at http://pubs.acs.org.

***References***

---

* To whom correspondence should be addressed. (R. C.) Fax: (+33) 5-56-84-56-00. E-mail: clerac@crpp-bordeaux.cnrs.fr.

† Centre de Recherche Paul Pascal

‡ Laboratoire Louis Néel

§ Tokyo Metropolitan University, CREST and PRESTO

*f* Present address. Faculty of Chemistry, Northeast Normal University, Changchun, 130024, China.

(1) (a) Wernsdorfer, W.; Sessoli, R. *Science* **1999**, *284*, 133; (b) Leuenberger, M. N.; Loss, D. *Nature* **2001**, *410*, 789.

(2) Jamet, M.; Wernsdorfer, W.; Thirion, C.; Mailly, D.; Dupuis, V.; Mélinon, P.; Pérez, A. *Phys. Rev. Lett*. **2001**, *86*, 4676.

(3) (a) Christou, G.; Gatteschi, D.; Hendrickson, D. N.; Sessoli, R. *MRS Bull.* **2000**, *25*, 66; (b) Gatteschi, D.; Sessoli, R. *Angew. Chem. Int. Ed.* **2003**, *42*, 268.

(4) (a) Caneschi, A.; Gatteschi, D.; Lalioti, N.; Sangregorio, C.; Sessoli, R.; Venturi, G.; Vindigni, A.; Rettori, A.; Pini, M. G.; Novak, M. A. *Angew. Chem. Int. Ed.* **2001**, *40*, 1760; (b) Caneschi, A.; Gatteschi, D.; Lalioti, N.; Sangregorio, C.; Sessoli, R.; Venturi, G.; Vindigni, A.; Rettori, A.; Pini, M. G.; Novak, M. A. *Europhys. Lett.* **2002**, *58*, 771; (c) Pardo, E.; Ruiz-Garcia, R.; Lloret, F.; Faus, J.; Julve, M.; Journaux, Y.; Delgado, F.; Ruiz-Perez, C. *Adv. Mater.* **2004**, *16*, 1597; (d) Sun, Z.-M.; Prosvirin, A. V.; Zhao, H.-H.; Mao, J.-G.; Dunbar, K. R. *J. Appl. Phys.* **2005**, *97*, 10B305; (e) Kajiwara, T.; Nakano, M.; Kaneko, Y.; Takaishi, S.; Ito, T.; Yamashita M.; Igashira-Kamiyama, A.; Nojiri, H.; Ono, Y.; Kojima, N. *J. Am. Chem. Soc.* **2005**, *127*, 10150.

(5) Clérac, R.; Miyasaka, H.; Yamashita, M.; Coulon, C. *J. Am. Chem. Soc.* **2002**, *124*, 12837.

(6) (a) Chang, F.; Gao, S.; Sun, H.-L.; Hou, Y.-L.; Su, G. *Proceeding of the ICSM **2002** conference, Shanghai, China* p.182 ; (b) Lescouëzec, R.; Vaissermann, J.; Ruiz-Pérez, C.; Lloret, F.; Carrasco, R.;

Julve, M.; Verdaguer, M.; Dromzee, Y.; Gatteschi, D.; Wernsdorfer, W. *Angew. Chem. Int. Ed.* **2003**, *42*, 1483; (c) Toma, L. M.; Lescouëzec, R.; Lloret, F.; Julve, M.; Vaissermann, J.; Verdaguer, M. *Chem. Commun.* **2003**, 1850; (d) Miyasaka, H.; Clérac, R.; Mizushima, K.; Sugiara, K.; Yamashita, M.; Wernsdorfer, W.; Coulon, C. *Inorg. Chem.* **2003**, *42*, 8203; (e) Liu, T.-F.; Fu, D.; Gao, S.; Zhang, Y.-Z.; Sun, H.-L.; Su, G.; Liu, Y.-J. *J. Am. Chem. Soc.* **2003**, 125, 13976; (f) Shaikh, N.; Panja, A.; Goswami, S.; Banerjee, P.; Vojtysek, P.; Zhang, Y.-Z.; Su, G.; Gao, S. *Inorg. Chem.* **2004**, *43*, 849; (g) Chakov, N. E.; Wernsdorfer, W.; Abboud, K. A.; Christou, G. *Inorg. Chem.* **2004**, *43*, 5919

(7) (a) Miyasaka, H.; Nezu, T.; Sugimoto, K.; Sugiara, K.; Yamashita, M.; Clérac, R. *Chem. Eur. J.* **2005**, *11*, 1592; (b) Ferbinteanu, M.; Miyasaka, H.; Wernsdorfer, W.; Nakata, K.; Sugiura, K.; Yamashita, M.; Coulon, C.; Clérac, R. *J. Am. Chem. Soc.* **2005**, *127*, 390.

(8) (a) Yoo, J.; Yamaguchi, A.; Nakano, M.; Krystek, J.; Streib, W. E.; Brunel, L. C.; Hishimoto, H.; Christou, G.; Hendrickson, D.N. *Inorg. Chem.* **2001**, *40*, 4604; (b) Yang, E. C.; Harden, N.; Werndorfer, W.; Zakhrov, L.; Brechin, E. K.; Rheingold, A. L.; Christou G.; Hendrickson, D. N. *Polyhedron* **2003**, *22*, 1857; (c) Hendrickson, D. N.; Christou, G.; Ishimoto, H.; Yoo, J.; Brechin, E. K.; Yamaguchi, A.; Rumberger, E. M.; Aubin, S. M. J.; Sun Z.; Aromí, G. *Polyhedron* **2001**, *20*, 1479.

(9) (a) Lecren, L.; Li, Y-G.; Wernsdorfer, W.; Roubeau, O.; Miyasaka, H.; Clérac, R. *Inorg. Chem. Commun*. **2005**, *8*, 626 ; (b) Lecren, L.; Wernsdorfer, W.; Li, Y-G.; Roubeau, O.; Miyasaka, H.; Clérac, R. *J. Am. Chem. Soc*. **2005**, *127,* 11311.

(10) Miyasaka, H.; Nakata, K.; Sugiara, K.; Yamashita, M.; Clérac, R. *Angew. Chem. Int. Ed.* **2004**, *43*, 707.

(11) Yoo, J.; Wernsdorfer, W.; Yang, E-C.; Nakano, M.; Rheingold, A. L.; Hendrickson, D. N. *Inorg. Chem.* **2005**, 44, 3377.

(12) Boudreaux, E. A.; Mulay, L. N. *Theory and Applications of Molecular Paramagnetism*, Eds. John Wiley & Sons, New-York, **1976**.

(13) Wernsdorfer, W. *Adv. Chem. Phys.* **2001**, *118*, 99.

(14) Otwinowski, Z.; Minor, W. *Methods Enzymol.* **1996**, *276*, 307.

(15) (a) Sheldrick, G. M. *SHELXL97*, Program for Crystal Structure Refinement; University of Göttingen: Göttingen, Germany, **1997**; (b) Sheldrick, G. M. *SHELXS97*, Program for Crystal Structure Solution; University of Göttingen: Göttingen, Germany, **1997**.

(16) (a) Brechin, E. K.; Yoo, J.; Nakano, M.; Huffman, J. C.; Hendrickson, D. N.; Christou, G. *Chem. Commun.* **1999**, *17*, 783; (b) Wittick, L. M.; Murray, K. S.; Moubaraki, B.; Batten, S. R.; Spiccia, L.; Berry, K. J. *Chem. Soc. Dalton Trans.* **2004**, 1003.

(17) Borrás-Almenar, J. J.; Clemente-Juan, J. M.; Coronado, E.; Tsukerblat, B. S. *J. Comput. Chem.* **2001**, *22*, 985.

(18) (a) Kennedy, B. J.; Murray, K. S. *Inorg. Chem.* **1985**, *24* 1552; (b) Kennedy, B. J.; Murray, K. S. *Inorg. Chem.* **1985**, *24*, 1557; (c) Miyasaka, H.; Clérac, R.; Wernsdorfer, W.; Lecren, L.; Bonhomme, C.; Sugiura, K.; Yamashita, M. *Angew. Chem. Int. Ed.* **2004**, *43*, 2801.

(19) In order to take into account the inter-complex interaction, the following definition of the susceptibility has been used:

$$\chi = \frac{\chi_0}{1 - \dfrac{2zJ'}{Ng^2\mu_B^2}\chi_0}$$

where $\chi_0$ is the susceptibility of the non-interacting units, z the number of nearest neighbors and *J'* is the magnetic interactions between units. See for example: (a) Myers, B.E. ; Berger L.; Friedberg, S. *J. Appl. Phys.* **1969**, *40*, 1149; (b) O'Connor, C. J. *Prog. Inorg. Chem.* **1982**, *29*, 203.

(20) (a) Abu-Youssef, M. A. M.; Escuer, A.; Gatteschi, D.; Goher, M. A. S; Mautner, F. A.; Vicente, R. *Inorg. Chem.* **1999**, *38*, 5716; (b) Villanueva, M.; Mesa, J. L.; Urtiaga, M. K.; Cortés, R.; Lezama, L.; Arriortua, M. I.; Rojo, T. *Eur. J. Inorg. Chem.* **2001**, 1581.

(21) Note that a model taking into account all free parameters ($g$, $J_{bb}$, $J_{wb}$ and $J'$) has also been considered and leads to a non convergence of the fitting procedure. Moreover, it is also worth noting that the obtained $J'$ value is smaller, but of the same order as the $J_{wb}$ intra-complex interaction. This indicates that this approach is in the limit of the mean-field approximation and the obtained value must be considered with caution. Nevertheless, this value is confirmed by the results obtained with the one-dimensional model.

(22) The actual $Mn^{II}$-$Mn^{II}$ interaction $J_{Mn\text{-}Mn}$ can be estimated from $J'$ by considering that $J_{Mn\text{-}Mn}S_{Mn}^2 = J'S_T^2$ with $S_{Mn} = 2$ and $S_T = 9$.

(23) Note that further recrystallizations do not change significantly the amount of missing links.

(24) Matsubara, F.; Yoshimura, K.; Katsura, S. *Can. J. Phys.* **1973**, *51*, 1053.

(25) This expression is equivalent to the exact result of F. Matsubara et al (equation 3.19 in ref 24) when $|J'|$ is larger than $k_BT$. We have chosen this approximation to emphasize the exponential behavior of the parallel susceptibility.

(26) Note that similar values of $\chi_\perp$ have been estimated from the slope of the $\chi T$ vs. $T$ plot in inset of Figure 9.

(27) Coulon, C.; Clérac, R.; Lecren, L.; Wernsdorfer, W.; Miyasaka, H. *Phys. Rev. B* **2004**, *69*, 132408.

(28) From the low frequency value of the $\chi'$ offset identified to $2\chi_\perp/3$, $\chi_\perp$ can be estimated at about 0.9 and 1.2 $cm^3mol^{-1}$ for the as-synthesized and recrystallized samples, respectively. These value are significantly higher than the one deduced from the $\ln(\chi T)$ vs. $1/T$, but also probably more accurate as $\chi_\perp$ is estimated at only one temperature also corresponding to the lowest available temperature.

(29) Below 3 K, the parallel susceptibility can be considered as negligible in comparison to the perpendicular component of the magnetic susceptibility of the odd segments. Therefore, the magnetic susceptibility of the odd segments is the amplitude ($\chi_0$-$\chi_\perp$) of the in-phase susceptibility or can be deduced from the maximum of $\chi''$ which is equal to $(\chi_0-\chi_\perp)/2$ in a Debye model (where $\chi_0$ is the zero frequency and zero dc field magnetic susceptibility of the system).

(30) Suzuki, M.; Kubo, R. *J. Phys. Soc. Jpn* **1968**, *24*, 51.

(31) Luscombe, J. H.; Luban, M.; Reynolds, J. P. *Phys. Rev. E* **1996**, *53*, 5852.

(32) Dhar, D.; Barma, M. *J. Stat. Phys.* **1980**, *22*, 259.

(33 ) When the concentration of defects, $c$, is small, the probability for a given site to belong to a segment of size $n$ is $P_n \approx nc^2e^{-cn}$. The average magnetization at time $t$ is given by : $m(t) = \sum_{n=1}^{\infty} P_n m_n(t) \approx c^2 \sum_{n=1}^{\infty} n m_{n0} e^{-cn - t/\tau(n)}$, where $m_n(t)$ and $m_{n0}$ are the magnetization of a segment of size $n$ at time $t$ and at $t$ = 0 respectively. Using the steepest descent approximation, the argument of the exponential is first maximized to determine the dominant term of the sum and this single term is then kept to determine the corresponding time dependence of the magnetization. This calculation is exactly the same in the ferromagnetic and antiferromagnetic cases and the resulting time dependence of $m$ is the same.

(34) We have calculated the two components of the normalized susceptibility using: $\chi'_{nor}(\omega) = \frac{1}{2\tau}\int_0^{\infty} \cos(\omega t)\exp(-\sqrt{t/\tau})dt$ and $\chi''_{nor}(\omega) = \frac{1}{2\tau}\int_0^{\infty} \sin(\omega t)\exp(-\sqrt{t/\tau})dt$.

# Supplementary material for the manuscript:

# Slow relaxation in a one-dimensional rational assembly of antiferromagnetically-coupled [$Mn_4$] single-molecule-magnets

*Lollita Lecren,*[†] *Olivier Roubeau,*[†] *Claude Coulon,*[†,]* *Yang-Guang Li,*[†,f] *Xavier Le Goff,*[†] *Wolfgang Wernsdorfer,*[‡] *Hitoshi Miyasaka,*[§] *and Rodolphe Clérac*[†,]*

Centre de Recherche Paul Pascal, CNRS UPR-8641, 115 av. Albert Schweitzer, 33600 Pessac, France; Laboratoire Louis Néel, CNRS, BP 166, 25 av. des Martyrs, 38042 Grenoble, France; Department of Chemistry, Graduate School of Science, Tokyo Metropolitan University, Minami-ohsawa 1-1, Hachioji, Tokyo 192-0397, PRESTO and CREST, Japan Science and Technology Agency, 4-1-8 Honcho, Kawaguchi, Saitama 332-0012, Japan.

E-mail: clerac@crpp-bordeaux.cnrs.fr; coulon@crpp-bordeaux.cnrs.fr

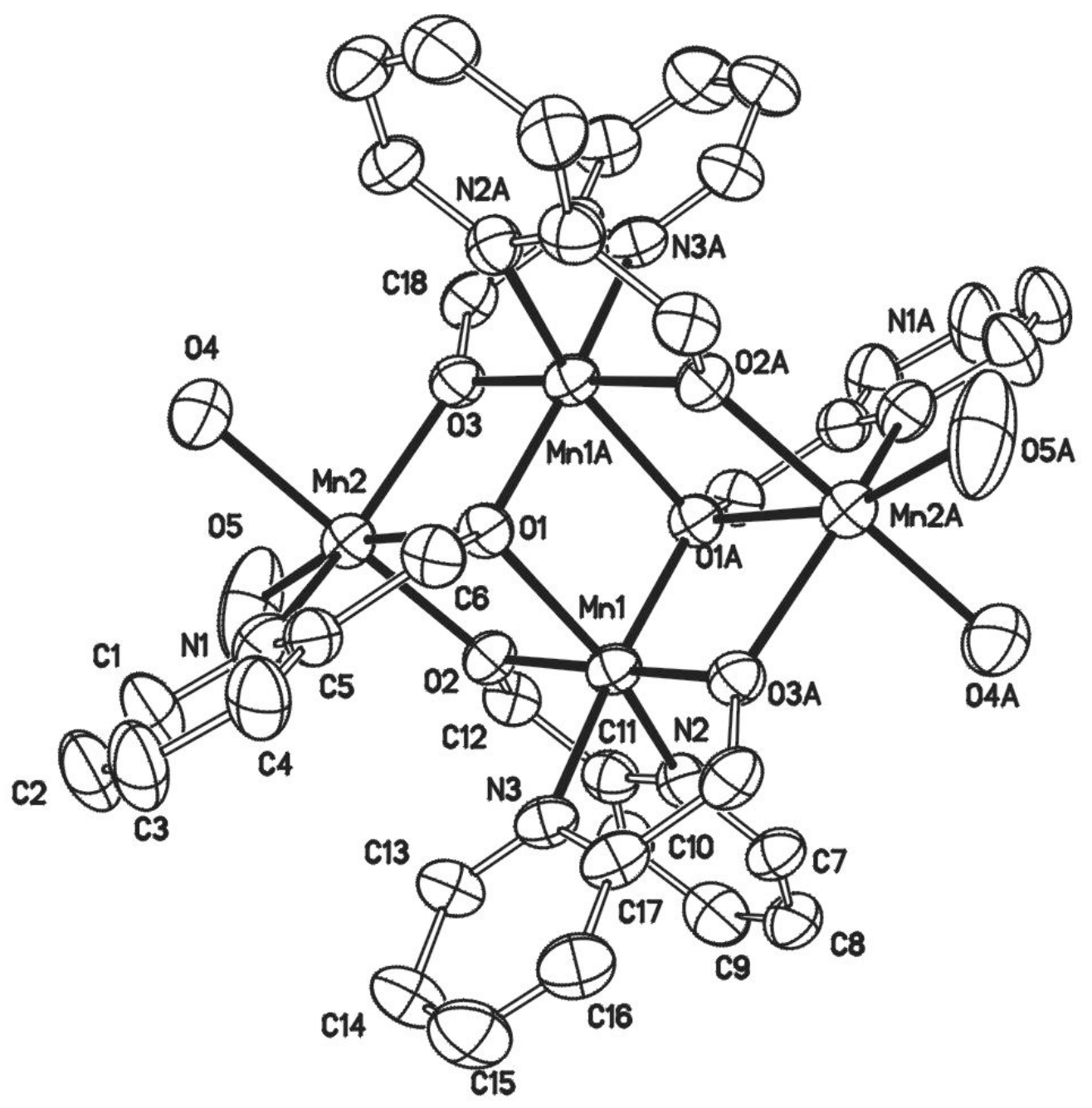

***Figure S1.*** ORTEP representation of the cationic part in **2** with thermal ellipsoids at 30 % probability. H atoms are omitted for clarity.

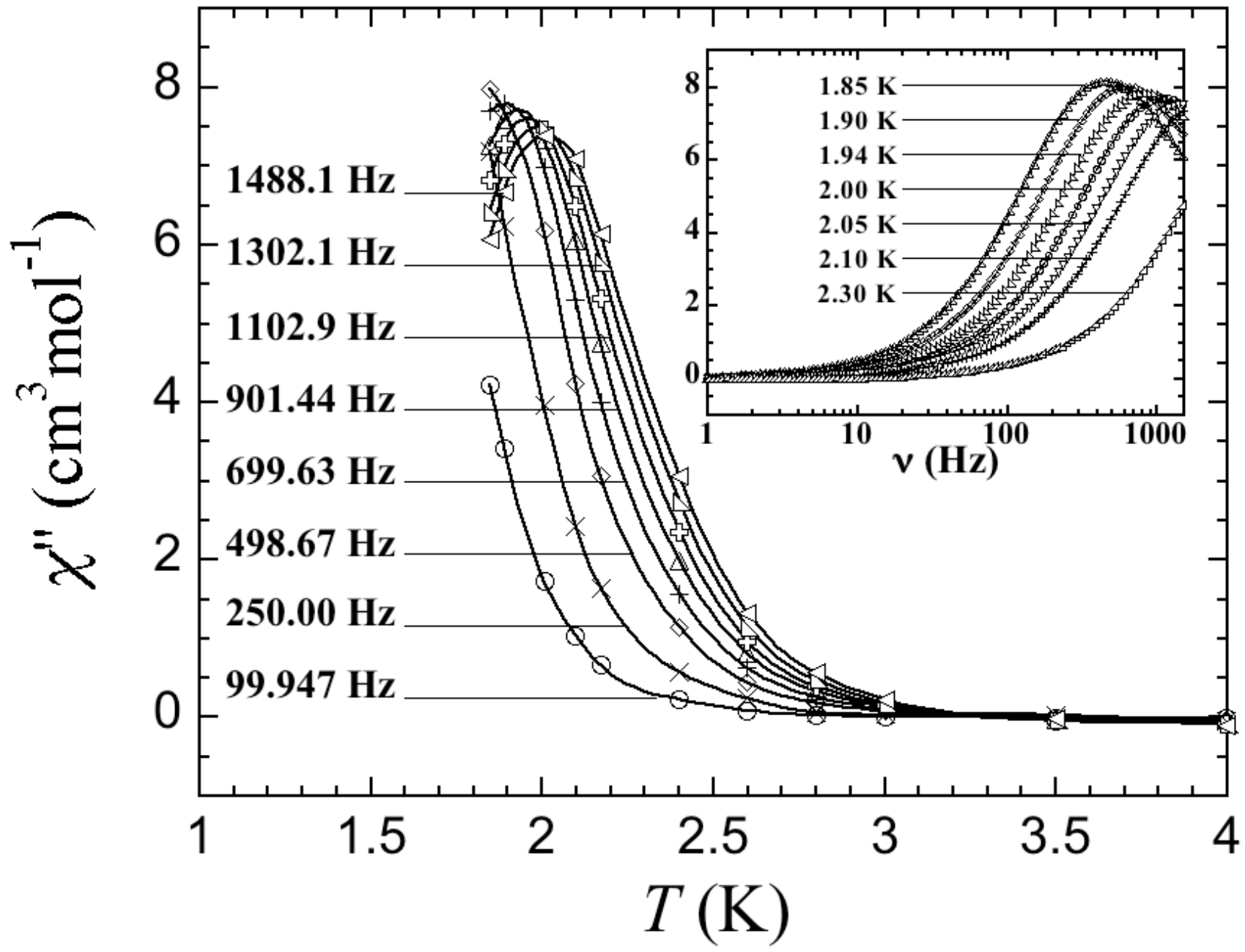

***Figure S2.*** Plot of $\chi$'' vs. $T$ under zero dc field for **1**. Inset: Plot of $\chi$'' vs. $\nu$ under zero dc field. Solid lines are only guides. (ref 9a)

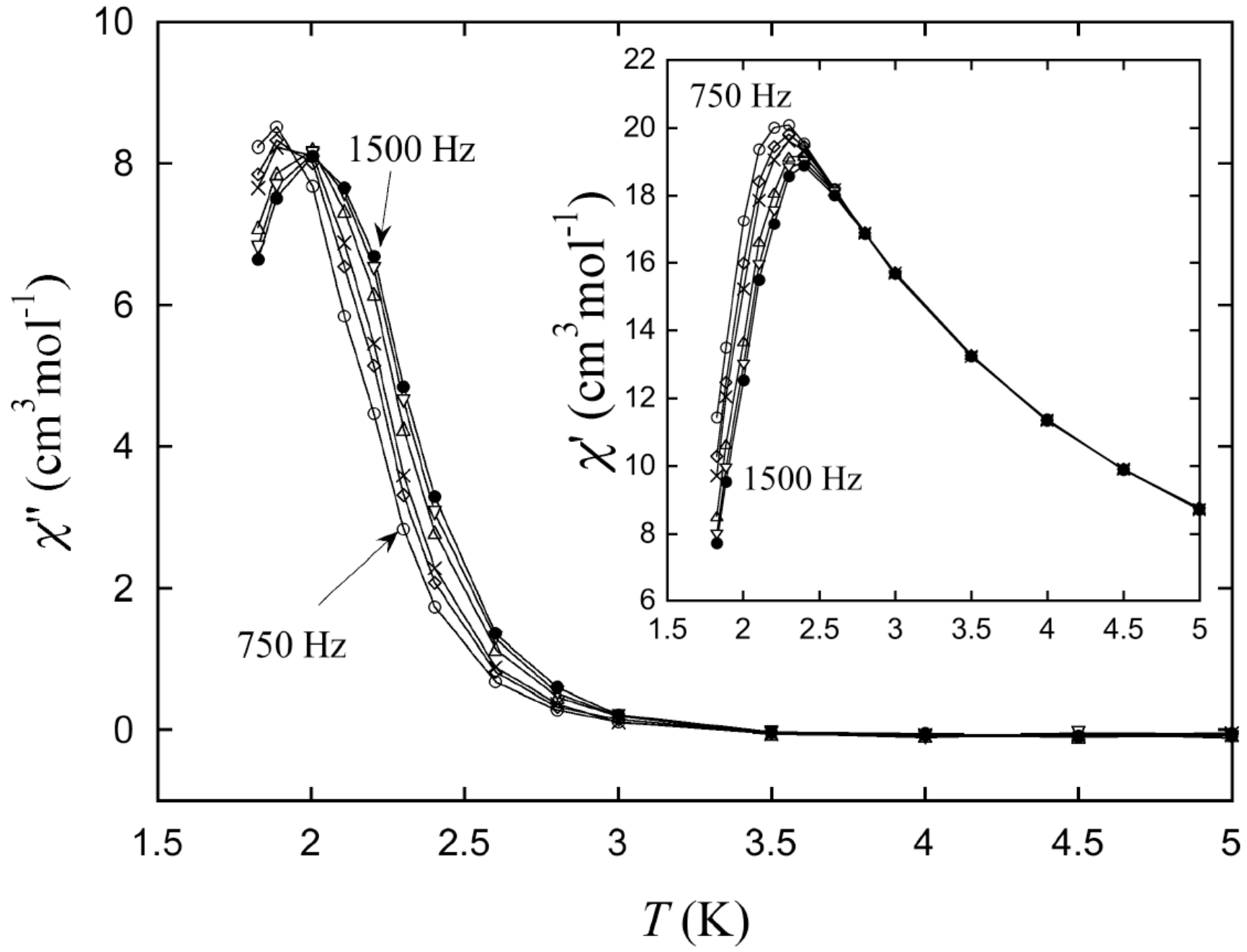

***Figure S3.*** Plot of $\chi$'' vs. $T$ under zero dc field for **2**. Inset: Plot of $\chi$' vs. $T$ under zero dc field. Solid lines are only guides.

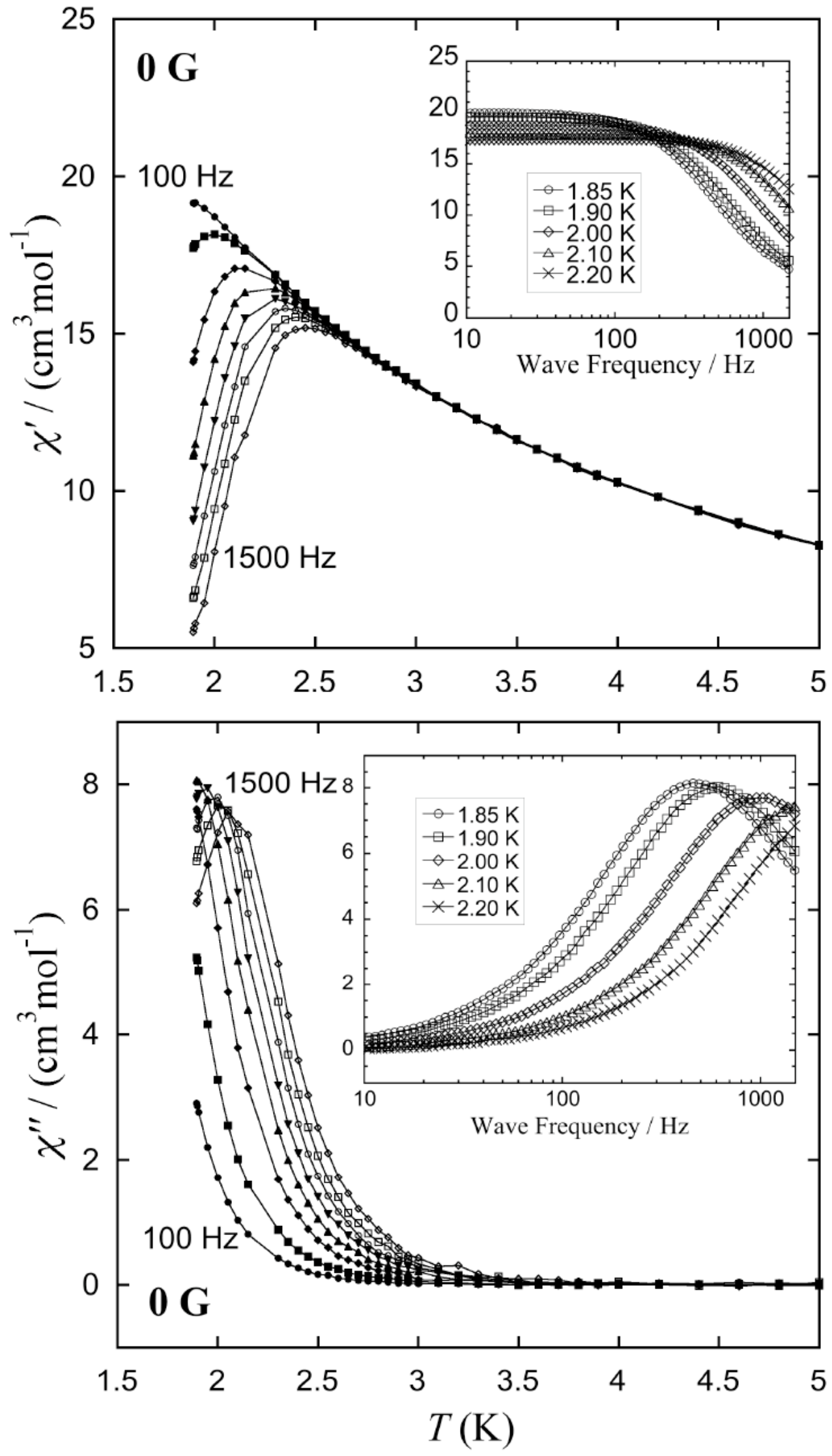

***Figure S4.*** Plot of $\chi$' (top) and $\chi$'' (bottom) vs. $T$ under zero dc field for **3**. Insets: Plot of $\chi$' (top) and $\chi$'' (bottom) vs. $\nu$ under zero dc field. Solid lines are only guides.

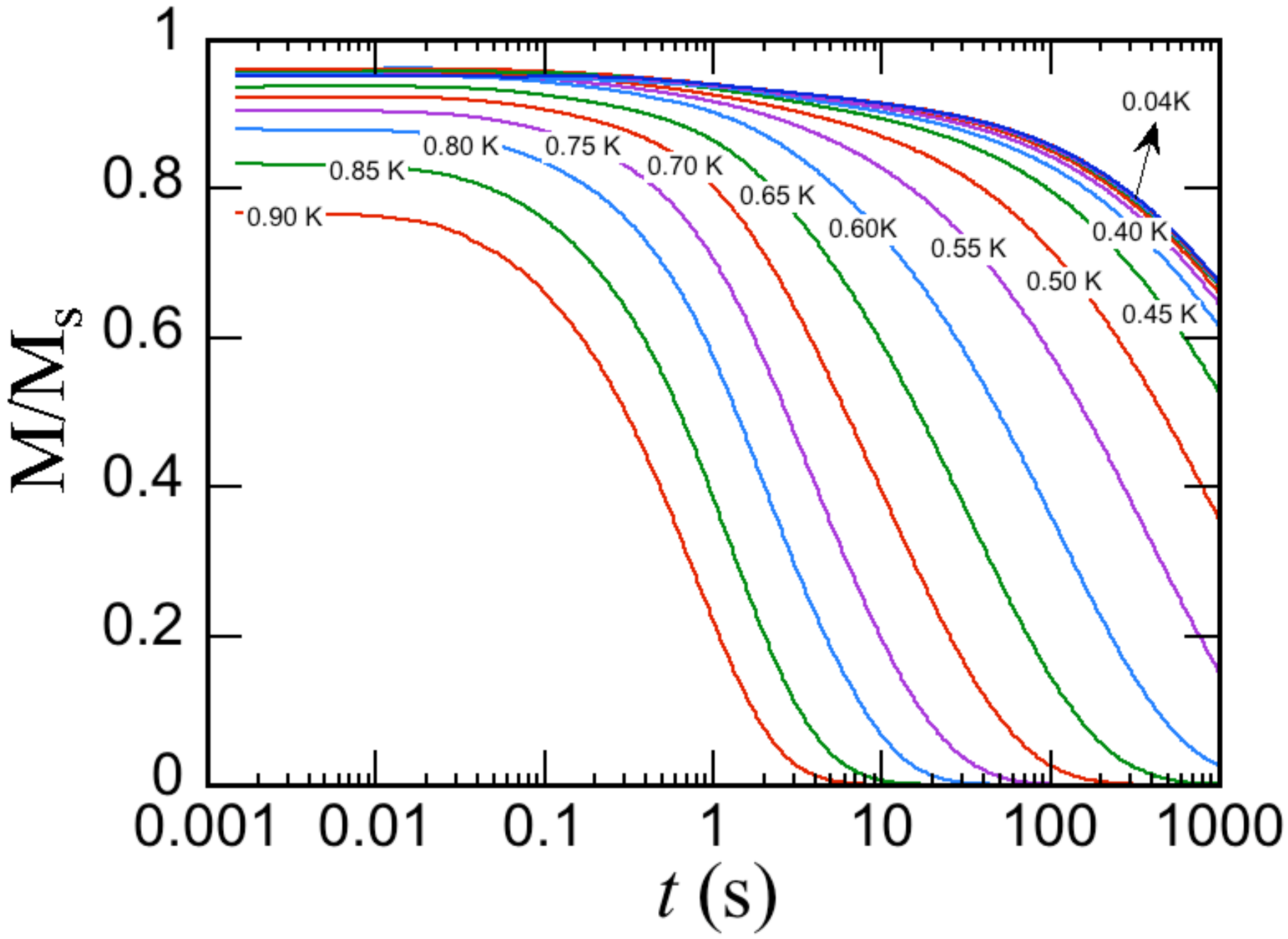

***Figure S5.*** Relaxation of the magnetization of **1** at different temperatures. The data are normalized to the saturation magnetization ($M_s$) at 1.4 T. (ref 9a)

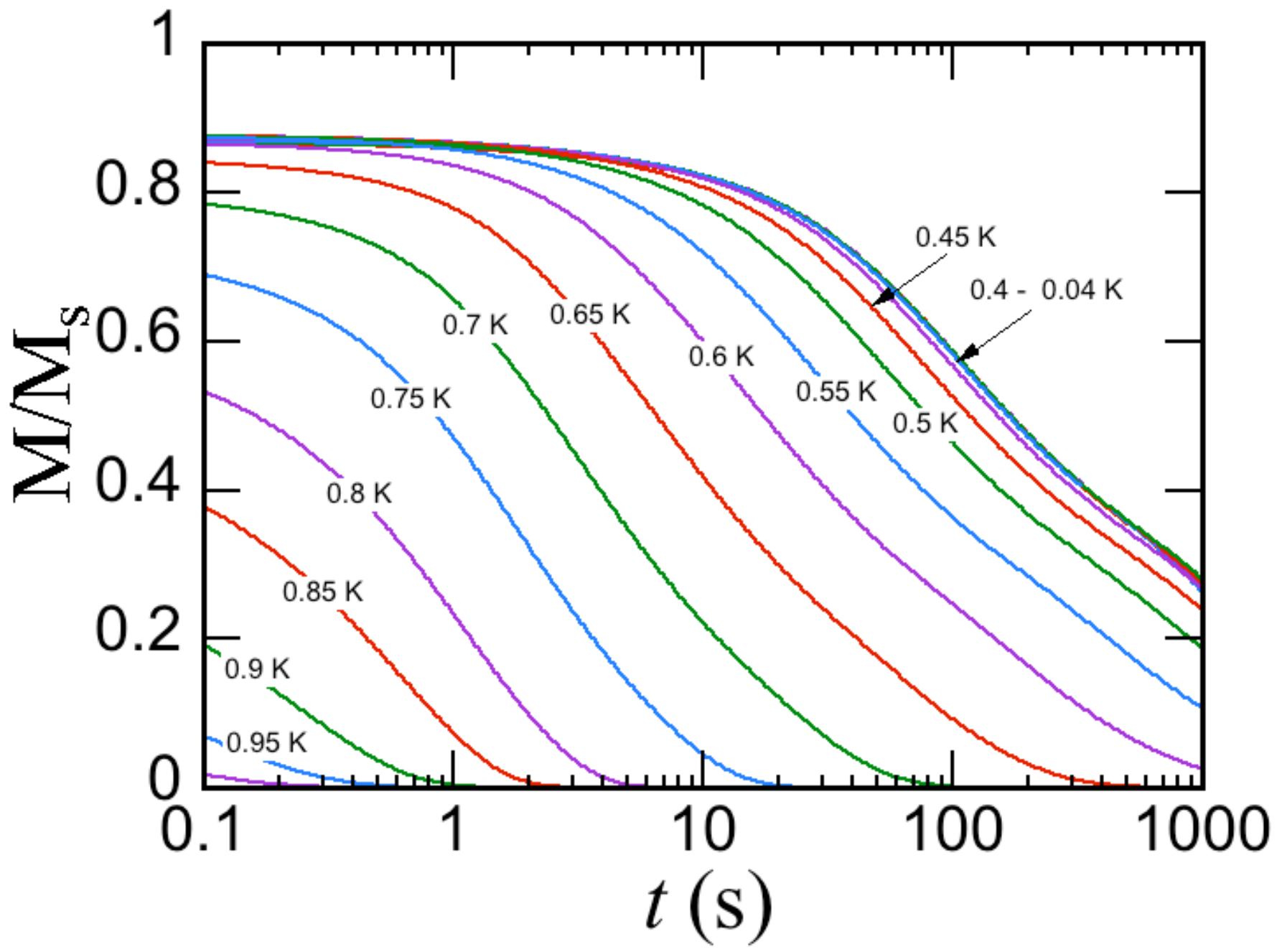

***Figure S6.*** Relaxation of the magnetization of **3** at different temperatures. The data are normalized to the saturation magnetization ($M_s$) at 1.4 T.

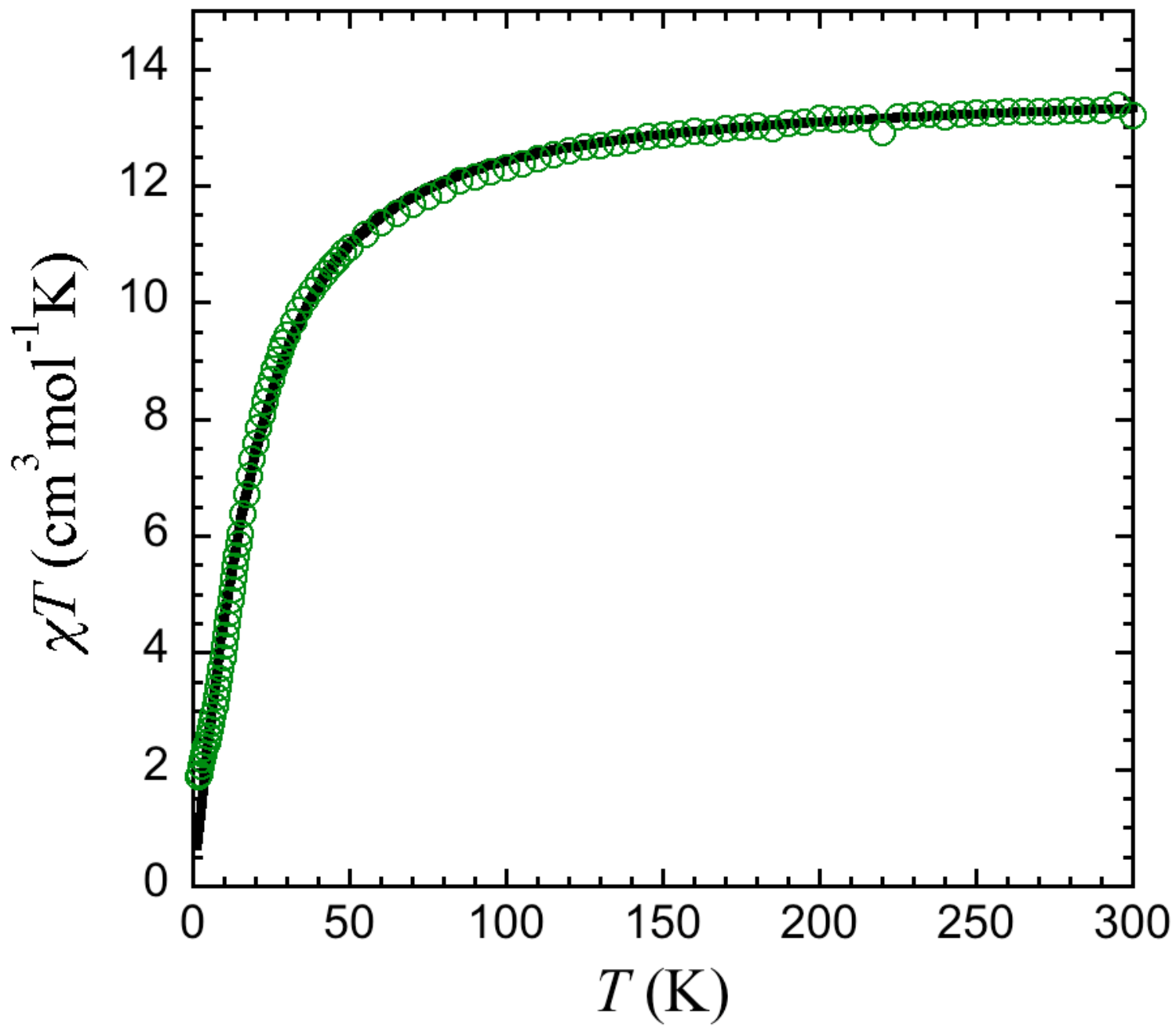

***Figure S7***. Plot of $\chi T$ *vs.* $T$ under 0.1 T for **5** (O). The solid lines represent the best fit obtained with the tetranuclear model described in the text taking into account antiferromagnetic inter-tetramer interactions within the chain in the mean-field approximation. Best-fit parameters: $g$ = 1.91(1), $zJ'/k_B$ = –0.31(4) K with $J_{wb}/k_B$ and $J_{bb}/k_B$ fixed respectively at +0.7 and +8.6 K. Note that convergence is not reached if all parameters are left free.